\documentclass[%
 reprint,
 amsmath,amssymb,
 aps,
]{revtex4-2}

\usepackage{graphicx}
\usepackage{dcolumn}
\usepackage{bm}
\usepackage [dvipsnames] {xcolor}
\usepackage{hyperref}
\hypersetup
{
colorlinks=true,
linkcolor=blue,
filecolor=blue,
urlcolor=blue,
citecolor=blue,
}
\usepackage[mathlines]{lineno}

\begin{document}

\preprint{APS/123-QED}

\title{Time Delay Anomalies of Fuzzy Gravitational Lenses}

\author{Jianxiang Liu$^{1}$}
\author{Zijun Gao$^{1}$}
\author{Marek Biesiada$^{2}$} 
\author{Kai Liao$^{1}$}%
 \email{E-mail:liaokai@whu.edu.cn}

\affiliation{$^{1}$Department of Astronomy, School of Physics and Technology, Wuhan University, Wuhan 430072, China\\
$^{2}$National Centre for Nuclear Research, Pasteura 7, PL-02-093 Warsaw, Poland
}

\date{\today}

\begin{abstract}
Fuzzy dark matter is a promising alternative to the standard cold dark matter. It has quite recently been noticed, that they can not only successfully explain the large-scale structure in the Universe, but can also solve problems of position and flux anomalies in galaxy strong lensing systems.
In this paper we focus on the perturbation of time delays in strong lensing systems caused by fuzzy dark matter, thus making an important extension of previous works. We select a specific system HS 0810+2554 for the study of time delay anomalies. Then, based on the nature of the fuzzy dark matter fluctuations, we obtain theoretical relationship between the magnitude of the perturbation caused by fuzzy dark matter, its content in the galaxy, and its de Broglie wavelength $\lambda _{\mathrm{dB}}$. It turns out that, the perturbation of strong lensing time delays due to fuzzy dark matter quantified as standard deviation is $\sigma_{\Delta t} \propto \lambda _{\mathrm{dB}}^{1.5}$. We also verify our results through simulations. Relatively strong fuzzy dark matter fluctuations in a lensing galaxy make it possible to to destroy the topological structure of the lens system and change the arrival order between the saddle point and the minimum
point in the time delays surface. Finally, we 
stress the unique opportunity for studying properties of fuzzy dark matter created by possible precise time delay measurements from strongly lensed transients like fast radio bursts, supernovae or gravitational wave signals. 
\end{abstract}

\maketitle


\section{Introduction}

The nature of the dark matter (DM) -- the dominant component of virialized objects: galaxies and their clusters --
is one of the biggest open questions in modern physics and astrophysics.
Especially the collision-less cold DM (CDM), which is widely recognized by the astrophysics community and strongly
supported by the current observations of large-scale structures (e.g. galaxy clusters), became a paradigm of modern cosmology as a part of the $\Lambda$CDM model \citep{10.1093/mnras/stx3304}. However, this paradigm still suffers from the well-known small-scale problem that concerns some observed features of galaxies and dwarf galaxies (e.g., missing satellite, cusp-core, and too-big-to-fail problems) \citep{bullock2017small}.   
Some studies attributed the small-scale problem to the poor modeling of baryonic processes and galaxy formation on small scales, rather than the CDM paradigm \citep{chan2015impact}.
There are two main conjectures about the nature of DM. One is Weakly Interacting Massive Particles (WIMPs) that have rest-mass energies $\gtrsim 10\;\mathrm{eV}$ \cite{jungman1996supersymmetric}. 

The other extreme of the mass scale, opposite to WIMPs is occupied by axions. Axions in a broad sense are a large class of bosons whose rest-mass energies $\ll 1\;\mathrm{eV}$ \cite{arvanitaki2010string,svrcek2006axions,marsh2016axion,hu2000fuzzy}.
Cosmological simulations treating them as cold dark matter also confirm expected rich nonlinear structure \cite{schive2014cosmic}. 
For a given halo mass, $M_{h}$ the de Broglie scale, $\lambda _{\mathrm{dB}}$, is set by the boson mass, $m_{\psi }$ according to the relationship \cite{schive2014understanding}: 
\begin{equation}
\label{equation:lambda}
    \lambda _{\mathrm{dB}}=150\left (\frac{10^{-22}\;{\mathrm{eV}}}{m_{\psi }}\right )\left ( \frac{M_{h}}{10^{12}M_\odot} \right )^{-1/3}{\mathrm{pc}}
\end{equation}

Considering that the occupancy number of axions in a volume of the de Broglie scale size is sufficiently large, the axion population can be well described by classical waves \cite{hui2021wave}. 
That is, self-interfering waves can modulate the density of the entire dark matter halo composed of axions on the de Broglie scale \cite{mocz2019first,woo2009high,schive2014cosmic,bullock2017small}.
Constraining the mass range of axions is a recent issue of significant concern.
Axions considered in this paper are ultralight axions whose masses are $m_{\psi} \sim 10^{-22}\;{\mathrm{eV}}$, also known as fuzzy dark matter \cite{hui2021wave}.
The $\lambda _\mathrm{dB}$ of ultralight axions varies from $10\;{\mathrm{pc}} \sim \mathrm{kpc}$ depending on the mass of given galaxy. 
As a result, the observable quantities of strong lensing system with such a lens galaxy may produce detectable perturbations caused by ultralight axions. 
Although some papers \citep{dalal2022excluding,nadler2021constraints,hui2021wave,schutz2020subhalo} in recent years suggested that the mass of axions might be $> 10^{-21}\;{\mathrm{eV}}$ , there still exists the possibility that axions partially contribute to the total components of DM halos. The study of perturbations caused by axions whose masses are $m_{\psi} \sim 10^{-22}\;{\mathrm{eV}}$ can help constrain the mass of axions using observational data from strong lensing systems.
In order to distinguish these two DM candidates and highlight their characteristics, we will denote WIMPs as $\rho \mathrm{DM}$ and axions as $\psi \mathrm{DM}$.
The fluctuation range of $\psi \mathrm{DM}$ density on a scale of $\lambda _\mathrm{dB}$ is between zero and twice the local mean density corresponding to destructive and constructive interference, respectively.  
Its two-dimensional density field projected on the plane perpendicular to the line of sight can be approximated as a Gaussian random field (GRF), see e.g. \cite{amruth2023einstein}. Such density fluctuations in galaxy halos can be observed through their effect on gravitational lensing.
Within the $\rho \mathrm{DM}$ scenario smoothly varying density profiles are sometimes unable to fully reproduce observed brightness and location of multiple images created by strong gravitational lensing  \cite{nierenberg2020double,keeton2003identifying,kochanek2004tests,goldberg2010fold,shajib2019every,xu2015well,hartley2019strong,spingola2018sharp,biggs2004radio,chan2020multiple}. They are known as brightness (or flux) anomaly and position anomaly, respectively. The flux  anomaly is usually tried to be explained by adding subhaloes to the galaxy's halo. This seems consistent with the long standing problem of too much power on the smallest scales revealed in $\rho \mathrm{DM}$ based N-body simulations. 
In some strong lensing systems, position and brightness anomalies can be accounted for by a better modeling of the lens, which includes dwarf galaxies or other perturbers \cite{nierenberg2014detection,amara2006simulations}. $\psi \mathrm{DM}$ might be able to solve the problem of position and brightness anomalies simultaneously, without referring to various substructures.
This was recently demonstrated in \cite{chan2020multiple,amruth2023einstein}, in the case of a particular well measured strong lensing system HS 0810+2554. This is the system, which we will study further below and our focus will be on time delay anomalies caused by the $\psi \mathrm{DM}$.

Strong gravitational lensing provides us with a valuable tool for studying the mass distribution in distant galaxies, including substructures. 
The time delay between image $i$ and image $j$ produced by the strong gravitational lensing effect is \cite{treu2010strong}:
\begin{equation}
\label{equation:td}
    \Delta t_{i,j}=\frac{(1+z_{{\mathrm{L}}})D_{{\mathrm{S}}}D_{{\mathrm{L}}}}{cD_{{\mathrm{LS}}}}\Delta \phi_{i,j}
\end{equation}
with
\begin{equation}
\label{equation:Fermat potential difference}
    \Delta \phi_{i,j} =\left[ \frac{(\vec{x _i}-\vec{\beta} )^2}{2}-\hat{\Psi}(\vec{x _i})-\frac{(\vec{x _j}-\vec{\beta} )^2}{2}+\hat{\Psi}(\vec{x _j}) \right]
\end{equation}
where $c$ is the speed of light; $\vec{x _i}$ and $\vec{x _j}$ are image angular positions on the sky; $\vec{\beta}$ denotes the source angular position; 
$\hat{\Psi}$  is the lensing potential; $D_\mathrm{L}$, $D_\mathrm{S}$, and $D_\mathrm{LS}$ are angular diameter distances to the lensing galaxy (deflector) located at redshift $z_\mathrm{L}$, to the source located at redshift $z_\mathrm{S}$ and between them, respectively.

Unlike the magnification, time delays are not affected by stellar microlensing and dust extinction.
At the same time, uncertainties in the background cosmological model and the radial mass profile 
of the lens can have an impact on the time delays \cite{keeton2009new,schneider2013source}. Using time delays ratio can mitigate the influence of these factors.  

In this paper, we go further than \cite{chan2020multiple,amruth2023einstein} and study the perturbation of strong lensing time delays due to $\psi \mathrm{DM}$.  
In order to highlight the perturbing effect of fuzzy dark matter on strong lensing systems, we do not consider the influence of subhalos and other complex baryonic structures.  
In our study, we use the best fit results \cite{aghanim2020planck} obtained by Planck in 2018 ($H_0=67.7\mathrm{\;km/s/Mpc}$, $\Omega_\mathrm{m}=0.31$).

This paper is organized as follows. In Section \ref{Sec II} we present our methodology comprising the choice of the specific strong lensing system and modeling its mass density distribution (Section \ref{2.1}), introduction of the $\psi \mathrm{DM}$ model (Section \ref{2.2}) and theoretical predictions of the scaling behaviour of $\psi \mathrm{DM}$ induced perturbations of position, time delay and time delay ratio anomalies. In Section \ref{3} we present and discuss the results of simulations. Finally, conclusions and perspectives are given in Section \ref{4}. 

\section{Methods} \label{Sec II}

In this section we present the details of our methodology. In brief, it consists of the following steps. First, in Section \ref{2.1} we fit the data regarding image positions, to several possible smooth $\rho$DM models: the singular isothermal sphere (SIS), singular isothermal ellipsoid (SIE) model, power law (PL) model (all of them considered with and without the external shear) and finally the Navarro-Frenk-White (NFW) model. Then we chose two best fitted ones from the above mentioned set of models and perturb them with the $\psi$DM. Our approach of adding the perturbation is described in Section \ref{2.2}. Since our goal is to assess the perturbation of time delays in the system, in Section \ref{2.3} we present some theoretical consideration regarding this issue.

\subsection{Smooth lens model}\label{2.1}

Following \citep{hartley2019strong,amruth2023einstein}, we focus our attention on the specific system HS 0810+2554.
It consists of a foreground elliptical galaxy with a quadruply imaged, lensed background galaxy. The redshift of the foreground galaxy is $z_\mathrm{L}=0.89$, while the redshift of the source is $z_\mathrm{S}=1.51$ \citep{hartley2019strong}. 
The background galaxy is lensed by the foreground galaxy in such a way that each of its structural components produces four observable images. These components include a quasar (QSO) at optical/near-IR wavelengths observed by the Hubble Space Telescope (HST) \citep{kochanek1999results} and a pair of jets at radio wavelengths observed by the European VLBI Network (EVN) \citep{hartley2019strong}. 
Due to current uncertainties in registering optical and radio reference frames up to 10 mas, quadruply-lensed optical images and jets registered at the radio wavelengths cannot be used simultaneously to constrain the lens model. 
While the images at the radio wavelengths can provide eight position constraints, images at the optical wavelengths can only provide four position constraints. Moreover, the image position accuracy of radio wavelength is higher than that of the optical QSO \citep{amruth2023einstein}. So we model the lens using the eight constraints provided by the image positions at the radio wavelength. To be more specific, the data we used come from the Table 5 of \citep{hartley2019strong}. 

We considered several candidate models for the smooth $\rho \mathrm{DM}$ lens model: SIS, SIE, PL and NFW, all of them with and without external shear. For the purpose of model fitting we used the {\tt lenstronomy} package, which allowed us to analyze different lensing galaxy models and freely add various perturbations or additional baryon structures. The above mentioned models required different numbers of free parameters $\Theta$. We fitted the models to the observed locations of lensed radio images using the $\chi^2$ objective function defined as: 
\begin{equation}
\label{equation:error}
\chi^2=\sum_i^N\frac{\left[\mathbf{x}_i^\mathrm{obs}-\mathbf{x}_i(\Theta)\right]^2}{\sigma_i^2},
\end{equation}
where $\mathbf{x}_i^\mathrm{obs}$ is the location of the observed component and $\mathbf{x}_i(\Theta)$ is the location predicted by the model; $N=8$ is the number of image components used to constrain the model and $\sigma_i$ is the position uncertainty for each component. Minimization of the objective function was performed using the MCMC method with uniform priors on all parameters $\Theta$. 
During the model fitting, we found that there was a degeneracy between  power law-index $\gamma$ and ellipticity in an elliptical PL profile. Hence, we focused on the SIE model (equivalent to PL with $\gamma=2$) with and without external shear. 
In particular, we used 11 free parameters $\Theta$ to generate this model: source location $(RA, Dec)$ for two jets (4 parameters); lensing galaxy location $(GalaxyRA, GalaxyDec)$; lensing galaxy ellipticity and position angle; 
Einstein radius $\theta_E$; external shear magnitude and position angle $(GSM,GSA)$. MCMC sampling and fitting revealed that the external shear had almost no impact on the results of our fitting (the values of the objective function - $\chi^2_{min}$ of the best fitted results were very close). Moreover, the best fitted external shear magnitude was also very small ($GSM< 0.0001$). The same occurred when other models of the lens were fitted. Based on these findings we further considered the models with no external shear. 
This also advantageous for our main focus, which is the impact of $\psi \mathrm{DM}$ perturbations on the measurable quantities like SL time delays. 

\begin{figure}
	\centering
	\begin{minipage}{\linewidth}
		\centering
		\includegraphics[width=\linewidth]{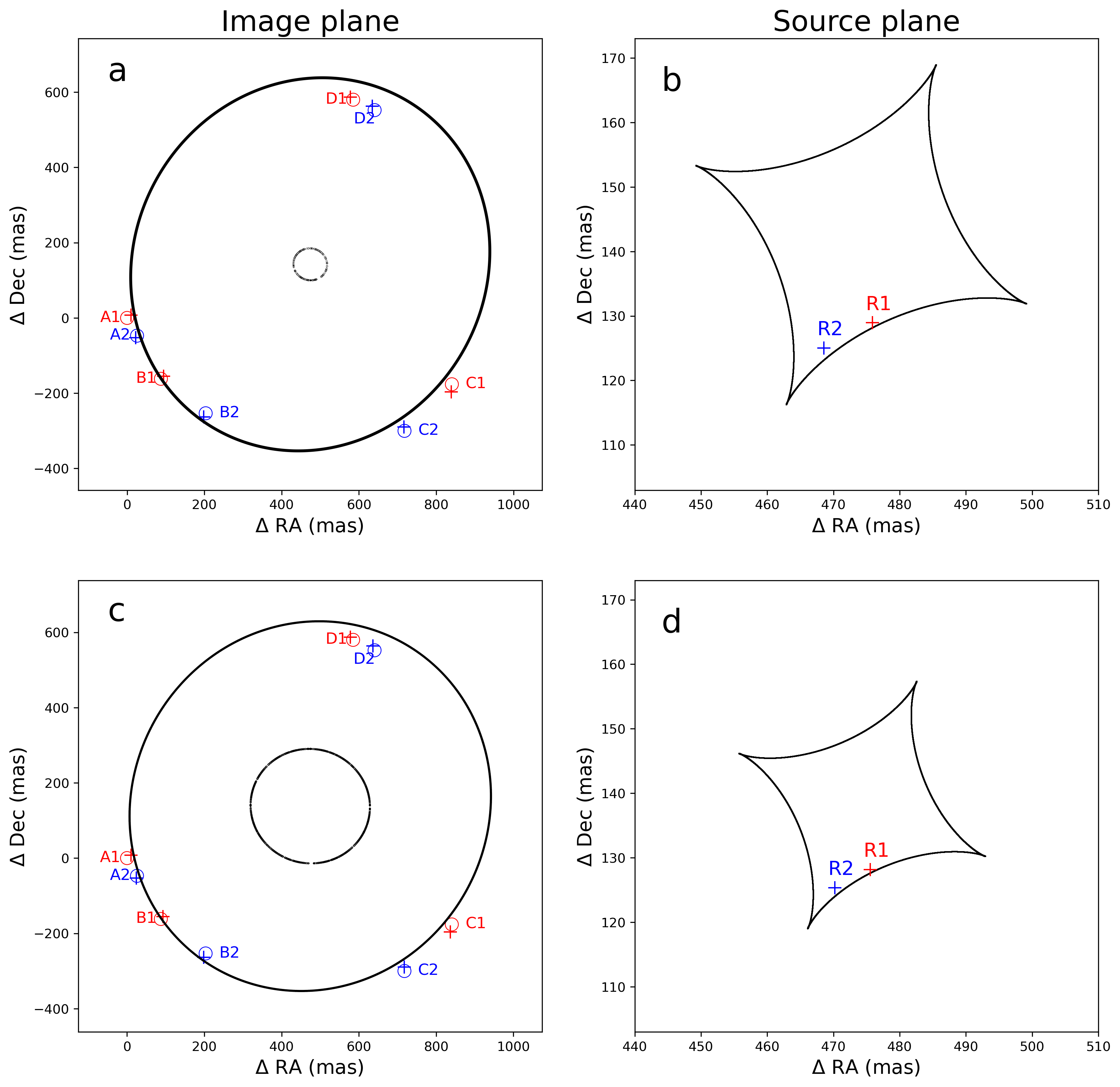}	
	\end{minipage}
    \caption{Schematic diagram of the location of the MCMC method’s best fit to the data. The results of elliptical PL profile are shown in a and b; the results of Navarro-Frenk-White (NFW) profile are shown in c and d. R1 and R2 in b and d are the best fitting positions of the two radio sources in the source plane, and the caustics of the corresponding models are also shown in b and d. The critical lines of the corresponding models are shown in a and c. A1-D2 respectively represent the eight images generated by the two radio sources, the open circles represent the observation results, and the crosses represent the fitting results.}
    \label{position.png}
\end{figure}

\begin{table}[b]
\centering
\caption{Parameters of the best fitted lens models used in our study. $\theta_E$ is the Einstein Radius and  $\alpha_\mathrm{R_\mathrm{s}}$ is the deflection angle at $R_\mathrm{s}$. Their values were derived from the MCMC best fit. }

\begin{tabular}{lcccc}
\hline
\hline
Model& $\theta_E$ (mas)&$\alpha_{R_\mathrm{s}} (\mathrm{mas})$&Ellipticity\\
\hline
PL($\gamma = 1.88$ )   & 479.3 &  - & 0.08\\
NFW($R_\mathrm{s}$= 626 mas)         &- &  495.5 & 0.02\\
\hline
\hline
\label{tab:para}
\end{tabular}
\end{table}

We found two best models with similar goodness of fit results. One is the elliptical PL profile. In the MCMC sampling, we fixed its lensing galaxy ellipticity to $\epsilon = 0.08$ which was the best fitted value of this parameter for the SIE model. Then we used 9 free parameters to generate an elliptical PL profile:  source location $(RA, Dec)$ for two jets (4 parameters);  
lensing galaxy location $(GalaxyRA, GalaxyDec)$; lensing galaxy position angle; Einstein radius $\theta_E$ and the power law index $\gamma$.
The second best fitted model was the Navarro-Frenk-White (NFW) profile. 
Considering that $\alpha_{R_\mathrm{s}}$ (deflection angle at '$R_\mathrm{s}$') is partly determined by $R_\mathrm{s}$, 
for the MCMC sampling, we fixed its scale radius $R_\mathrm{s}$ to $5\;\mathrm{kpc}$ \citep{amruth2023einstein} which is equivalent to  
$R_\mathrm{s} = 626 \; \mathrm{mas}$ in the angular units. Then we used 9 free parameters to generate a NFW model: source location $(RA, Dec)$ for two jets (4 parameters); lensing galaxy location $(GalaxyRA, GalaxyDec)$; lensing galaxy ellipticity and position angle; the deflection angle at $R_\mathrm{s}$ ($\alpha_{R_\mathrm{s}}$). 
As already discussed the models were tested with and without external shear, which turned out to be insignificant. Parameters of the best fitted models are summarized in Table \ref{tab:para}.
Fig. \ref{position.png} displays the best fitted models by showing the position of images on top of critical curves and caustics.

\begin{figure*}
    \centering
    \includegraphics[width=2\columnwidth]{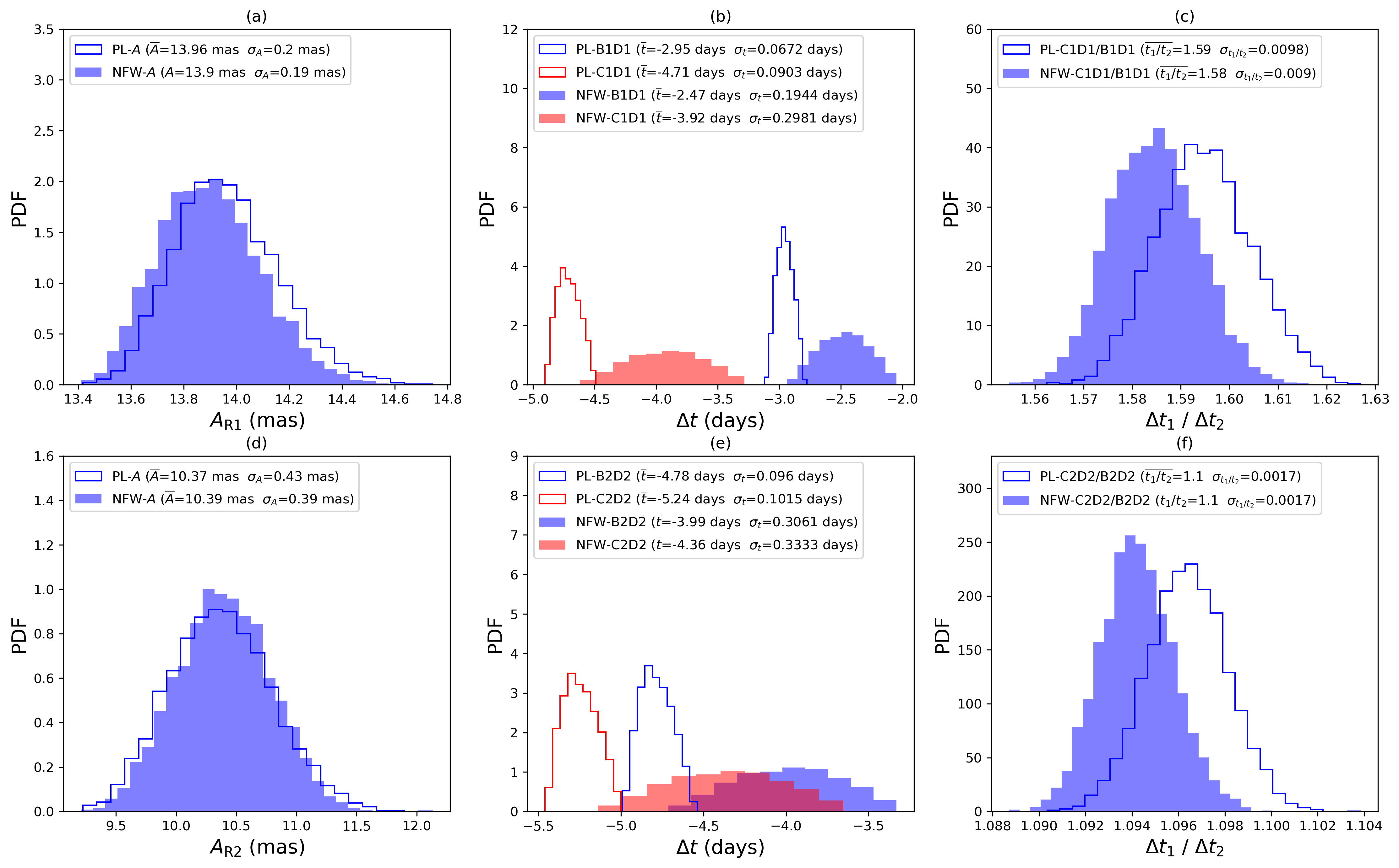}
    \caption{Histograms of the position anomaly $A$ - panels (a,d), time delays (b,e) and time delays ratios (c,f) obtained by the MCMC sampling within $1\sigma$ of smooth $\rho \mathrm{DM}$lens model. In the legends of the respective plots, PL and NFW represent the lens model, while B1,C1,B2,C2 represent the images considered. 
    }
    \label{pl_nfw_error_from_measurement.png}
\end{figure*}

\begin{figure}[b]
	\centering
	\begin{minipage}{\linewidth}
		\centering
		\includegraphics[width=\linewidth]{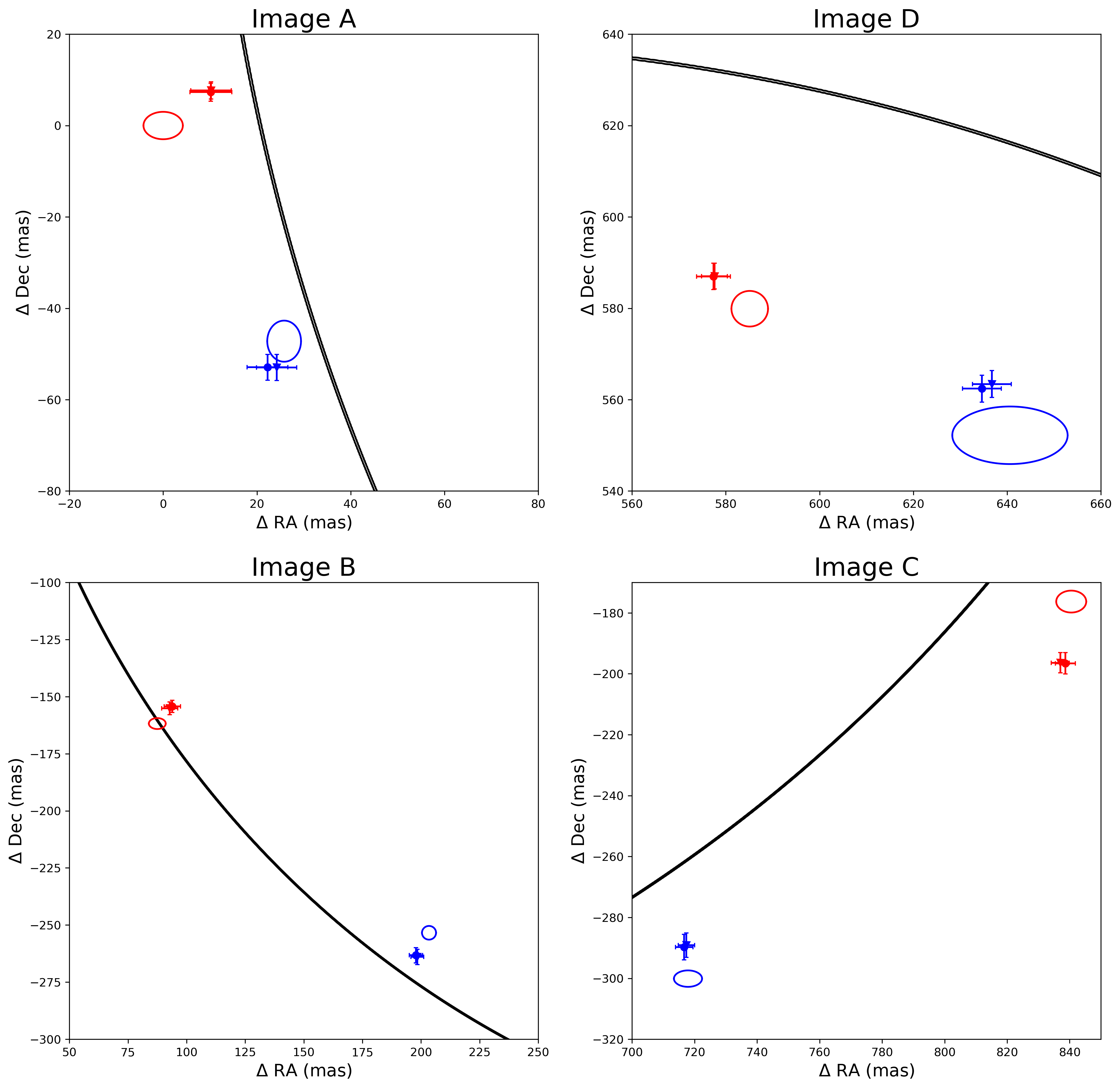}	
	\end{minipage}
    \caption{Position mismatch between predictions of the smooth $\rho \mathrm{DM}$ lens model and observed positions of images. Observed image locations are represented by ellipses, with the semi-major and semi-minor axes of each ellipse corresponding to the  $3\sigma$ location uncertainty ($\sigma$ is the measurement uncertainty) encompass $99.7\%$ of all possible positions. 
    Model predicted location of images is represented by uncertainty bar crosses, centered at the best fitted values. Circles represent elliptical PL model and inverted triangles correspond to NFW model. Sizes of the uncertainty bars correspond to $\pm3\sigma_\mathrm{tol}$, where $\sigma_\mathrm{tol}$ reflects the image position uncertainty coming from the model uncertainty}. 
    \label{nfw_pl_errorbar.png}
\end{figure}

We use two best models obtained above to represent the smooth $\rho \mathrm{DM}$ lens model.
To specify the error between the fitted model (smooth $\rho \mathrm{DM}$ lens model) and the observed image position data, we define the position anomaly as: $A=\left[\frac{1}{4}\sum_{i=1}^{4}(\mathbf{x}_{i}^{\varrho\mathrm{DM}} - \mathbf{x}_{i}^{\mathrm{observed}} ) ^ { 2 }\right]^{1/2}$.
Meanwhile, for the elliptical PL profile and NFW profile obtained above, we select the data within $1\sigma$ from the corresponding MCMC sampling. In other words, the selected data sets are all within one standard deviation of the corresponding parameter-sampled data. Then, we use these data to compute the corresponding position anomalies - $A$, time delays, time delay ratios, and $\sigma_\mathrm{tol}$. The position tolerance $ \sigma_\mathrm{tol}$ is the standard deviation of the predicted position computed for data within $1\sigma$ of the smooth $\rho \mathrm{DM}$ lens model, i.e. the uncertainty in the position prediction due to lens model uncertainty, not measurement uncertainty. The results of position anomalies-$A$, time delays and time delay ratios are displayed in Fig. \ref{pl_nfw_error_from_measurement.png}, while Fig. \ref{nfw_pl_errorbar.png} shows the comparison between $\sigma_\mathrm{tol}$ and the  measurement uncertainty. One can see in Fig. \ref{nfw_pl_errorbar.png}, that $\sigma_\mathrm{tol}$ is close to the magnitude of the measurement uncertainty, but the smooth $\rho \mathrm{DM}$ lens model does not explain the observed data, i.e. there are position anomalies in the lensing system HS 0810+2554.
In Fig. \ref{pl_nfw_error_from_measurement.png}, it can be seen that position anomalies - $A$ of the two models are very similar and both have an order of $\sim 10\;\mathrm{mas}$, which is similar to the results reported in the \citep{hartley2019strong}. 
At the same time, due to the presence of radial profile degeneracy, two best models have similar position predictions, but time delays have an overall scaling \citep{keeton2009new,schneider2013source}. However, the results for the time delay ratio are similar, i.e. the time delay ratio essentially eliminates this scaling. In other words, in order to eliminate this radial profile degeneracy, other measurements are needed to complement the modelling of the lens model. 
The reason we have chosen these two models is that they can be used to investigate whether the perturbation of the smooth lens by $\psi$DM to the time delay would be affected by this degeneracy.
Also, in Fig. \ref{position.png}, we can see that two jets $(R1,R2)$ we used for modelling are located at the fold and the cusp of the caustic, respectively. Hence, we can use the study of time delays of multiply-lensed two jets to analyse the perturbation of $\psi$DM to time delays of the source for two different positions in source plane.

\subsection{Adding the $\psi \mathrm{DM}$ perturbation}\label{2.2}

The column mass density along the line-of-sight direction coming from the $\psi \mathrm{DM}$ halo can be approximated by a Gaussian random field (GRF) of a characteristic scale $\lambda_\mathrm{dB}$ (Eq. (\ref{equation:lambda})) superimposed on the global density profile (i.e. a smooth $\rho$DM lens model obtained in Section \ref{2.1}) \citep{amruth2023einstein}.
In this section we are focused on the theoretical aspects of $\psi \mathrm{DM}$ perturbations. Therefore, for simplicity we neglect here the ellipticity of the lens model. Ellipticity will be included in the simulations presented and discussed later in Section \ref{3}. 
From the best fitted models, obtained in Section \ref{2.1}, one can estimate the halo mass $M_h \sim {10}^{12}M_\odot$ and by virtue of Eq. (\ref{equation:lambda}), the characteristic de Broglie scale as $\lambda_\mathrm{dB} \sim 180\;\mathrm{pc}$ (assuming $m_{\psi} \sim 10^{-22}\;{\mathrm{eV}}$). Given the analytic function of 
the global density profile of the $\rho \mathrm{DM}$ halo, the variance of its associated GRF $\sigma_\mathrm{\Sigma}$ is \cite{amruth2023einstein}:
\begin{equation}
\label{equation:sigma}
\begin{split}
\sigma_\mathrm{\Sigma}^2(\boldsymbol{\xi})&=\lambda_{\mathrm{dB}}\int_{-\infty}^\infty\sigma_\rho^2(z,\boldsymbol{\xi})dz\\& \simeq\lambda_{\mathrm{dB}}\int_{-\infty}^\infty\rho_{\mathrm{smooth}}^2(z,\boldsymbol{\xi})dz,
\end{split}
\end{equation}
where $z$ is the coordinate of the $z$-axis along the line-of-sight direction, $\boldsymbol{\xi}$ is the projection radius vector from the center of the halo, $\sigma_\rho$ is the standard deviation in the $\psi \mathrm{DM}$ halo density fluctuation at $(z,\boldsymbol{\xi})$, $\rho_\mathrm{smooth}$ is the global 3-D density profile. Approximate equality in the above formula is because $\sigma_\rho\sim \rho_{\mathrm{smooth}}(z,\boldsymbol{\xi})$ as the 3-D density fluctuates between zero and twice the local mean density. 

For the PL profile:
\begin{equation}
\label{equation:PL}
\begin{split}
\sigma_\mathrm{\Sigma}^2(\boldsymbol{\xi})\simeq2\rho_0^2\lambda_\mathrm{dB}\boldsymbol{\xi}^{-2\gamma+1}(4.34-5.03\gamma+2.84\gamma^2 \\ -0.77\gamma^3+0.08\gamma^4)
\end{split}
\end{equation}
$\gamma$ is the power law index of this profile, $\rho_0$ is 3-D density parameter of the PL profile. In the paper \citep{amruth2023einstein} an analogous formula for the NFW profile is provided.
Regarding the relation between $\sigma_\rho$ and $\rho_{\mathrm{smooth}}$ the specific quantitative relationship is not of great concern. 
Since the baryonic matter in the galaxy damps the perturbations from the dark matter, and due to the axially symmetric distribution of baryons, the damping is essentially distributed in an axially-symmetric manner.  
In our specific case, we are mainly concerned with the locations near the critical line of the system under study. Following the arguments presented in \cite{amruth2023einstein}, the damping introduced by smoothly-distributed baryons decreases away from the halo centre. So it is reasonable to consider damping of the surface mass density fluctuations by a constant factor. 

\begin{figure}
	\centering
	\begin{minipage}{\linewidth}
		\centering
		\includegraphics[width=\linewidth]{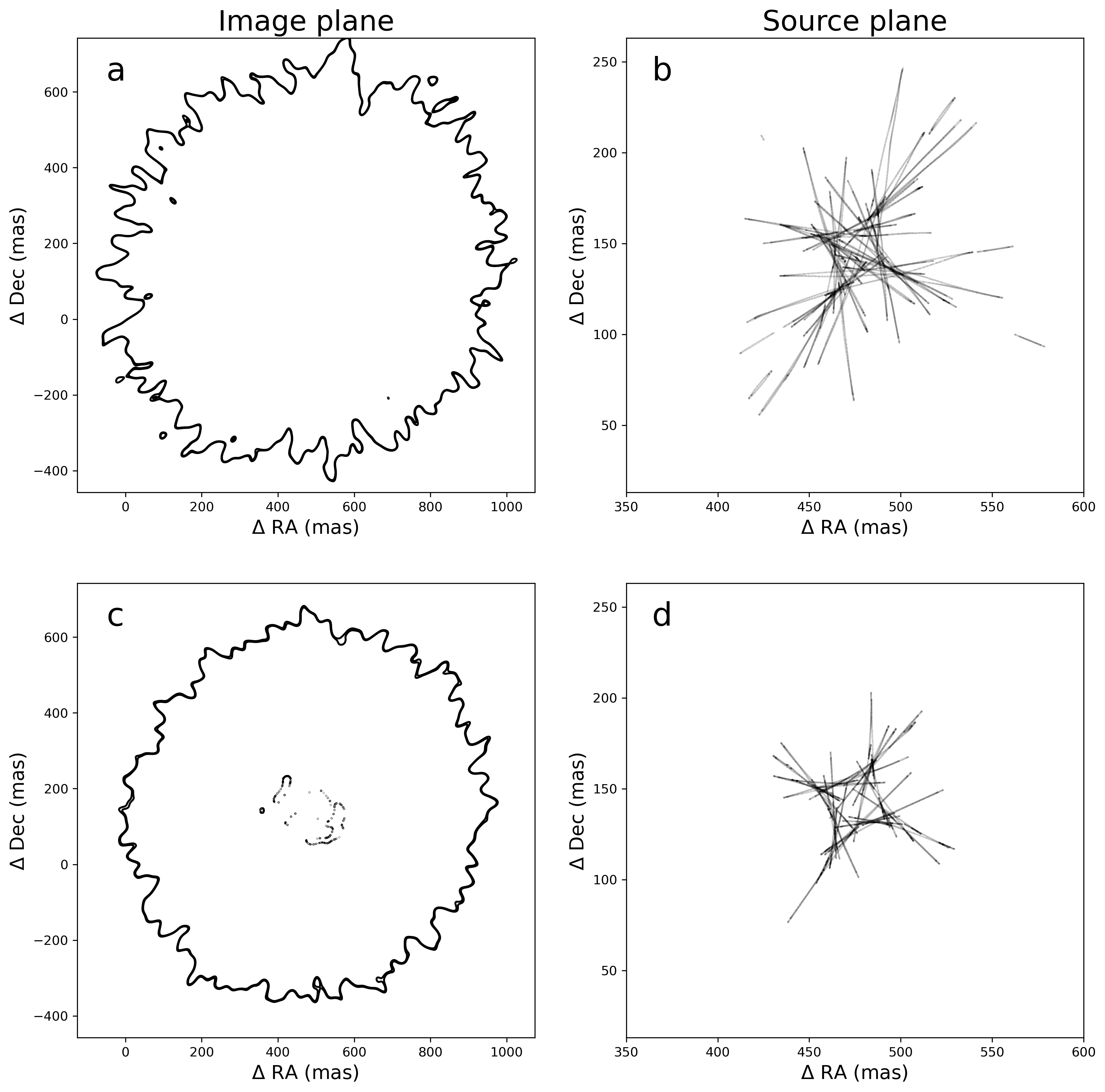}	
	\end{minipage}
    \caption{Critical lines and caustics of $\psi \mathrm{DM}$ halo assuming $\lambda_\mathrm{dB} = 180\;\mathrm{pc}$, superposed over the best fitted elliptical PL profile. Proportionality coefficient is $d = 1$ in panels (a,b) and $d = 0.5$ in (c,d). }
    \label{critical line perturbation.png}
\end{figure}

One can assume a simple proportionality $\sigma_\rho(z,\boldsymbol{\xi})=d \, \rho_{\mathrm{smooth}}(z,\boldsymbol{\xi})$, where $d$ is a constant coefficient. Different magnitude of the $\psi \mathrm{DM}$ perturbation of the smooth global profile will have differently pronounced impact on the critical lines and caustics, which is illustrated in Fig. \ref{critical line perturbation.png}.

\begin{figure}
	\centering
	\begin{minipage}{\linewidth}
		\centering
		\includegraphics[width=\linewidth]{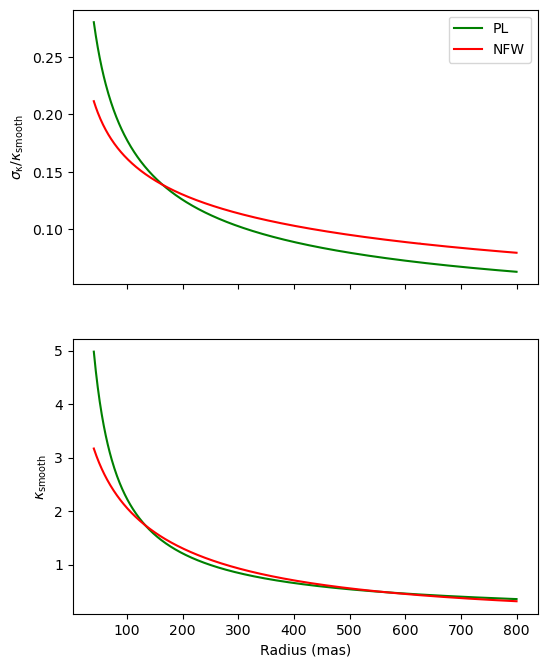}	
	\end{minipage}
    \caption{Plot showing how the relative perturbation $\sigma_\kappa(\boldsymbol{\xi})/\kappa_{\mathrm{smooth}}$ of the smooth model by $\psi$DM and the convergence $\kappa_{\mathrm{smooth}}$ change with the radius $\boldsymbol{\xi}$ in PL (green line) and NFW (red line) models.}
    \label{perturbation intensity.png}
\end{figure}

In the theory of gravitational lensing one usually introduces a dimensionless quantity $\kappa$, called convergence:
\begin{equation}
\label{equation:rhocr}
\kappa(\boldsymbol{\xi})\equiv\frac{\Sigma(\boldsymbol{\xi})}{\Sigma_{\mathrm{cr}}}
\end{equation}
where $\Sigma(\boldsymbol{\xi})$ is the surface mass density of the lensing galaxy and 
 $ \Sigma_{\mathrm{cr}}=\frac{c^2}{4\pi G}\frac{D_\mathrm{S}}{D_\mathrm{L}D_\mathrm{LS}} $
is called the critical surface density, which is a function of the angular diameter distances of lens and source. 
Therefore, the $\sigma_\Sigma^2(\boldsymbol{\xi})$ can also be expressed in dimensionless version: $\sigma_\kappa^2(\boldsymbol{\xi})\equiv\sigma_\Sigma^2(\boldsymbol{\xi})/\Sigma_\mathrm{cr}^2$.

It can be seen in Fig. \ref{perturbation intensity.png}, that the relative magnitude of the $\psi \mathrm{DM}$ perturbation expressed as $\sigma_\kappa(\boldsymbol{\xi})/\kappa_{\mathrm{smooth}}$ decreases with increasing radius $\boldsymbol{\xi}$. Moreover, both the relative magnitude of the perturbation the surface density $\kappa_{\mathrm{smooth}}$ of the PL model and the NFW model are very close to each other, which may imply that the $\psi \mathrm{DM}$ perturbation may be similar in both models. This suggests that the magnitude of the $\psi \mathrm{DM}$ perturbation to time delays may not be affected by the radial profile degeneracy.  
In Section \ref{3.1} we will see that this is indeed the case. 

\subsection{Scaling of time delay perturbations with $\lambda_\mathrm{dB}$ - theoretical prediction}\label{2.3}

In this section, we analyze how density fluctuations affect position anomalies and time delays between images. In particular we are interested in how does the magnitude of the perturbation of time delays, substantiated as the variance of time delays, scale with the de Broglie length $\lambda_\mathrm{dB}$ of $\psi \mathrm{DM}$. 
Although we consider only axially symmetric lenses, the conclusions obtained in this section can be directly generalized to more general models. 

To begin with, let us take $x _{i0}$ as image angular locations, $y_0$ as a source angular location without $\psi \mathrm{DM}$ perturbation and $x _i$ as image angular locations with $\psi \mathrm{DM}$ perturbation. All the above mentioned quantities are dimensionless, i.e. expressed in units of the Einstein radius $\theta_E$. Similarly, it would be convenient to introduce a dimensionless de Broglie length $\hat{\lambda}_\mathrm{dB} = \lambda_\mathrm{dB} / R_E$, where $R_E = D_L \theta_E$ is physical Einstein radius in $[\mathrm{pc}]$. 
According to the theory of gravitational lensing, one can write the lens equation:
\begin{equation}
\begin{aligned}
&y_0 = x_{i0} - \alpha_0(x_{i0}) \\
&y_0 = x_i - \alpha(x_i)
\end{aligned}
\label{eq:without and with}
\end{equation}
where $\alpha_0(x _{i0})$ and $\alpha(x _{i})$ are the dimensionless deflection angles without and with perturbation, respectively. Combining Eq. (\ref{eq:without and with}) , and using the approximations: 
$\alpha_0(x _{i})-\alpha_0(x _{i0})=\dot{\alpha_0}(x _{i0})(x _{i}-x _{i0})+O(x _{i}-x _{i0})$ and $\alpha(x _{i})-\alpha_0(x _{i})\approx \alpha(x _{i0})-\alpha_0(x _{i0})$ one obtains:
\begin{equation}
\label{equation:positon perturbation2}
x _{i}-x _{i0}\approx\frac{\alpha(x _{i0})-\alpha_0(x _{i0})}{1-\dot{\alpha_0}(x _{i0})}
\end{equation}

Applying the well known formula:
\begin{equation}
\label{equation:alpha}
\alpha(x)=\frac{2}{x}{\int_{0}^{x}\kappa(x')x'\mathrm{d}x'}
\end{equation}
where $\kappa$ is the dimensionless two-dimensional surface density, one can express the variance of position mismatch as:
\begin{equation}
\label{equation:positon perturbation3}
\begin{split}
\sigma^2(x _{i}-x _{i0})&\approx\frac{\sigma^2(\int_{0}^{x_{i0}}(\kappa(x')-\kappa_0(x'))x'\mathrm{d}x')}{x_{i0}^2(1-\dot{\alpha_0}(x _{i0}))^2}\\& =\frac{\sigma^2(\int_{0}^{x_{i0}}\kappa(x')x'\mathrm{d}x')}{x_{i0}^2(1-\dot{\alpha_0}(x _{i0}))^2}
\end{split}
\end{equation}

According to the Generalized Central Limit Theorem (Lindeberg's condition), if we are dealing with independent but differently distributed random variables, then the standard deviation (or variance) of the sum of such variables is still limited by the standard normal distribution when certain conditions are met \citep{lindeberg1922neue,CLT}. Scales of the $\psi$DM relevant to our study are $\lambda_\mathrm{dB} \sim 30 - 270\; \mathrm{pc}$, which is much smaller than the physical Einstein radius scale of order of few $\mathrm{kpc}$. And we has $\hat{\lambda}_\mathrm{dB} \ll 1$.
Hence, we can think of a small segment $\Delta x$ of length $\hat{\lambda}_\mathrm{dB}$ as an independent part and approximate the integral in the numerator of Eq. (\ref{equation:positon perturbation3}) in the following way:

\begin{equation}
\label{equation:positon perturbation4}
\begin{split}
&\sigma^2(\int_{0}^{x}\kappa(x)x\mathrm{d}x)\approx\Delta x^2\sigma^2(\sum_{i=0}x_i\kappa(x_i))\\& \approx \hat{\lambda}_\mathrm{dB}^2\sigma^2(\sum_{i=0}x_i\kappa(x_i))
=\hat{\lambda}_\mathrm{dB}^2\sum_{i=0}\sigma^2(x_i\kappa(x_i))\\& \approx \hat{\lambda}_\mathrm{dB}^2\sum_{i=0}\frac{\int_{x_i}^{x_i+\hat{\lambda}_\mathrm{dB}}x'^2\sigma^2(\kappa(x'))\mathrm{d}x'}{\hat{\lambda}_\mathrm{dB}}\\& =\hat{\lambda}_\mathrm{dB} \int_{0}^{x_{i0}} x'^2\sigma^2(\kappa(x'))\mathrm{d}x'
\end{split}
\end{equation}
Eq. (\ref{equation:sigma}) implies that $\sigma^2(\kappa(x))\propto\hat{\lambda}_\mathrm{dB}$, so $\sigma^2(x _{i}-x _{i0})\propto{\hat{\lambda}_\mathrm{dB}}^2$. Thus we have $\sigma(x_{i}-x _{i0})\propto\hat{\lambda}_\mathrm{dB}$ and using well known properties of the variance, one can see that $\sigma((x _{i}-x _{i0})^2)\propto \hat{\lambda}_\mathrm{dB}$

Next we consider the variance of the perturbation of the lensing potential $\Psi(\vec{x})$ by $\psi \mathrm{DM}$.
\begin{equation}
\label{equation:potential perturbation1}
\Psi(\vec{x})=\int_{\mathbb{R}^2}\kappa(\vec{x^{\prime}})g(\vec{x}-\vec{x^{\prime}})\mathrm{d}^2x^{\prime}
\end{equation}
where
\begin{equation}
\label{equation:g}
g(\vec{x}-\vec{x}')=\frac1\pi\ln\left|\vec{x}-\vec{x}'\right|
\end{equation}
Similar to the previous calculations leading from Eq. (\ref{equation:positon perturbation2}) to Eq. (\ref{equation:positon perturbation4}), one can write:
\begin{equation}
\label{equation:potential perturbation2}
\begin{split}
&\sigma^2(\Psi(\vec{x})-\Psi_0(\vec{x_0}))\\&=\sigma^2(\Psi(\vec{x})-\Psi_0(\vec{x})+(\Psi_0(\vec{x})-\Psi_0(\vec{x_0}))
\\& \approx\sigma^2(\Psi(\vec{x})-\Psi_0(\vec{x}))+\sigma^2(\Psi_0(\vec{x})-\Psi_0(\vec{x_0})))
\end{split}
\end{equation}
Where $\Psi_0$ is the lensing potential without $\psi \mathrm{DM}$ perturbation and $\Psi$ is the lensing potential with $\psi \mathrm{DM}$ perturbation. Evaluating the first term, one arrives at the following scaling behavior:
\begin{equation}
\label{equation:potential perturbation3}
\begin{split}
&\sigma^2(\Psi(\vec{x})-\Psi_0(\vec{x}))\approx\sigma^2(\Psi(\vec{x_0})-\Psi_0(\vec{x_0}))
\\& =\sigma^2(\int_{\mathbb{R}^2}\kappa(\vec{x^{\prime}})g(\vec{x}-\vec{x^{\prime}})\mathrm{d}^2x^{\prime})
\\& \approx \hat{\lambda}_\mathrm{dB}^4\sum_{i,j}\sigma^2(\kappa(\vec{x_{i,j}^{\prime}})g(\vec{x}-\vec{x_{i,j}^{\prime}}))
\\& \approx\hat{\lambda}_\mathrm{dB}^4\frac{\sum_{i,j}\int_{\Delta\mathbb{R}^2}\sigma^2(\kappa(\vec{x_{i,j}^{\prime}}))g^2(\vec{x}-\vec{x_{i,j}^{\prime}})\mathrm{d}^2 x^{\prime}}{\hat{\lambda}_\mathrm{dB}^2}\propto\hat{\lambda}_\mathrm{dB}^3
\end{split}
\end{equation}
where the final scaling is due to $\sum_{i,j}\int_{\Delta\mathbb{R}^2}\sigma^2(\kappa(\vec{x_{i,j}^{\prime}}))g^2(\vec{x}-\vec{x_{i,j}^{\prime}})\mathrm{d}^2 x^{\prime}\propto\hat{\lambda}_\mathrm{dB}$. Regarding the second term, one has:
\begin{equation}
\label{equation:potential perturbation4}
\begin{split}
\sigma^2(\Psi_0(\vec{x})-\Psi_0(\vec{x_0})))\sim \sigma^2(x _{i}-x _{i0})\propto\hat{\lambda}_\mathrm{dB}^2
\end{split}
\end{equation}

\begin{figure*}
    \centering
    \includegraphics[width=2\columnwidth]{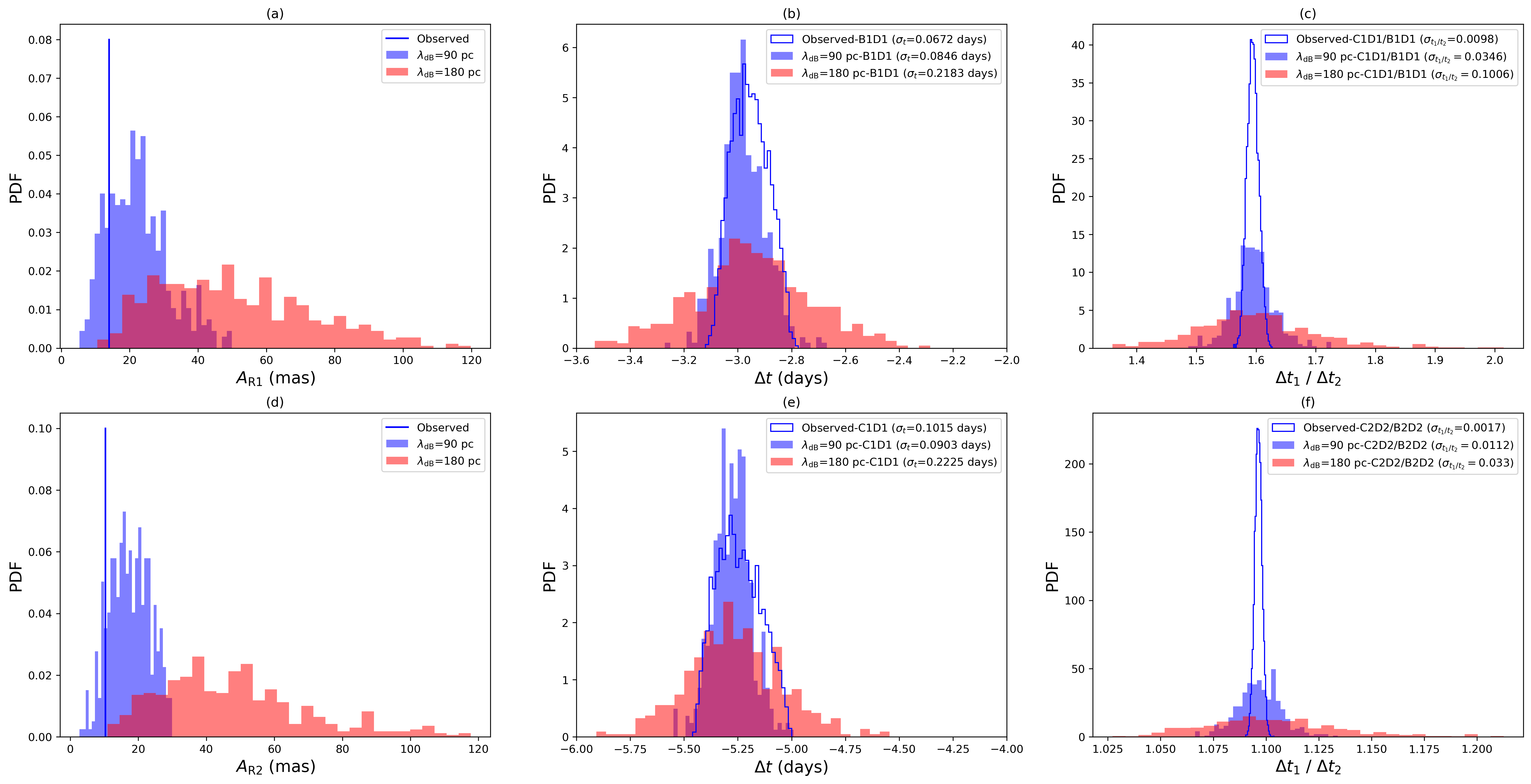}
    \caption{Histograms of the position anomaly - $A$, time delays and time delays ratios anomalies. In panels (a) and (d), 'Observed' indicates the position anomaly $A$ (defined in Section \ref{2.1}) of the best-fitting parameter in the MCMC sampling of the elliptical PL model, others anomalies indicate the position anomaly $\hat{A}$ (defined in Chapter \ref{3}) caused by $\psi \mathrm{DM}$ perturbation under different parameters of the elliptical PL model. In (b), (c), (e) and (f), the data of 'Observed' are the same as the data of the corresponding elliptical PL model in Fig. \ref{pl_nfw_error_from_measurement.png}; the rest of the data are similar with elliptical PL model in Fig. \ref{pl-nfw-compare.png}, but the specific parameters are different. In all cases reported strength of $\psi \mathrm{DM}$ perturbation is $d = 0.3$.}
    \label{change_lambda.png}
\end{figure*} 
Regarding perturbations of time delay $\sigma_{\Delta t}$, let us recall that by virtue of Eq. (\ref{equation:td}) and Eq. (\ref{equation:Fermat potential difference}) and using basic properties of the variance, one has:  $\sigma^2_{\Delta t} = \sigma^2((\vec{x _i}-\vec{y})^2) + \sigma(\hat{\Psi}(\vec{x _i}))$. Now, from the discussion following Eq. (\ref{equation:positon perturbation4}), we know that: $\sigma((\vec{x _i}-\vec{y})^2) \propto \hat{\lambda}_\mathrm{dB}$. Analogously, from the Eq. (\ref{equation:potential perturbation3}) one has: $\sigma(\hat{\Psi}(\vec{x _i})) \propto \lambda_\mathrm{dB}^{1.5}$. Hence, whenever $\lambda_\mathrm{dB}$ is not too small, the lens potential is the main factor affecting scaling of the variance of time delays and one has:  $\sigma_{\Delta t}\propto \lambda_\mathrm{dB}^{1.5}$ . According to the transfer characteristics of variance, one can immediately get $\sigma_{\Delta t_1/\Delta t_2}\propto \lambda_\mathrm{dB}^{1.5}$. Regarding proportionality coefficient $d$, it can be directly seen $\sigma_{\Delta t}\propto d$ and $\sigma({\theta _{i}-\theta _{i0}})\propto d$ based on the reasoning presented above.

We should emphasize that the above calculations were perturbative and neglected higher order terms.
Hence, they may not represent the real case accurately. In the next section we will test and verify these assessments in simulations.

\section{Simulation and discussion}\label{3}

For the purpose of simulations of the $\psi$DM perturbation added to the smooth lens potential, we used the {\tt powerbox} package \citep{murray2018powerbox} to generate the corresponding GRF. Then, the generated GRF were superimposed onto elliptical PL profile and NFW profile. In the specific calculations, we generated a GRF with a pixel size of $10\;\mathrm{pc}$ ($1.25\;\mathrm{mas}$) and spanning a rectangular area of 1000×1000 pixels.

\begin{figure*}
    \centering
    \includegraphics[width=2\columnwidth]{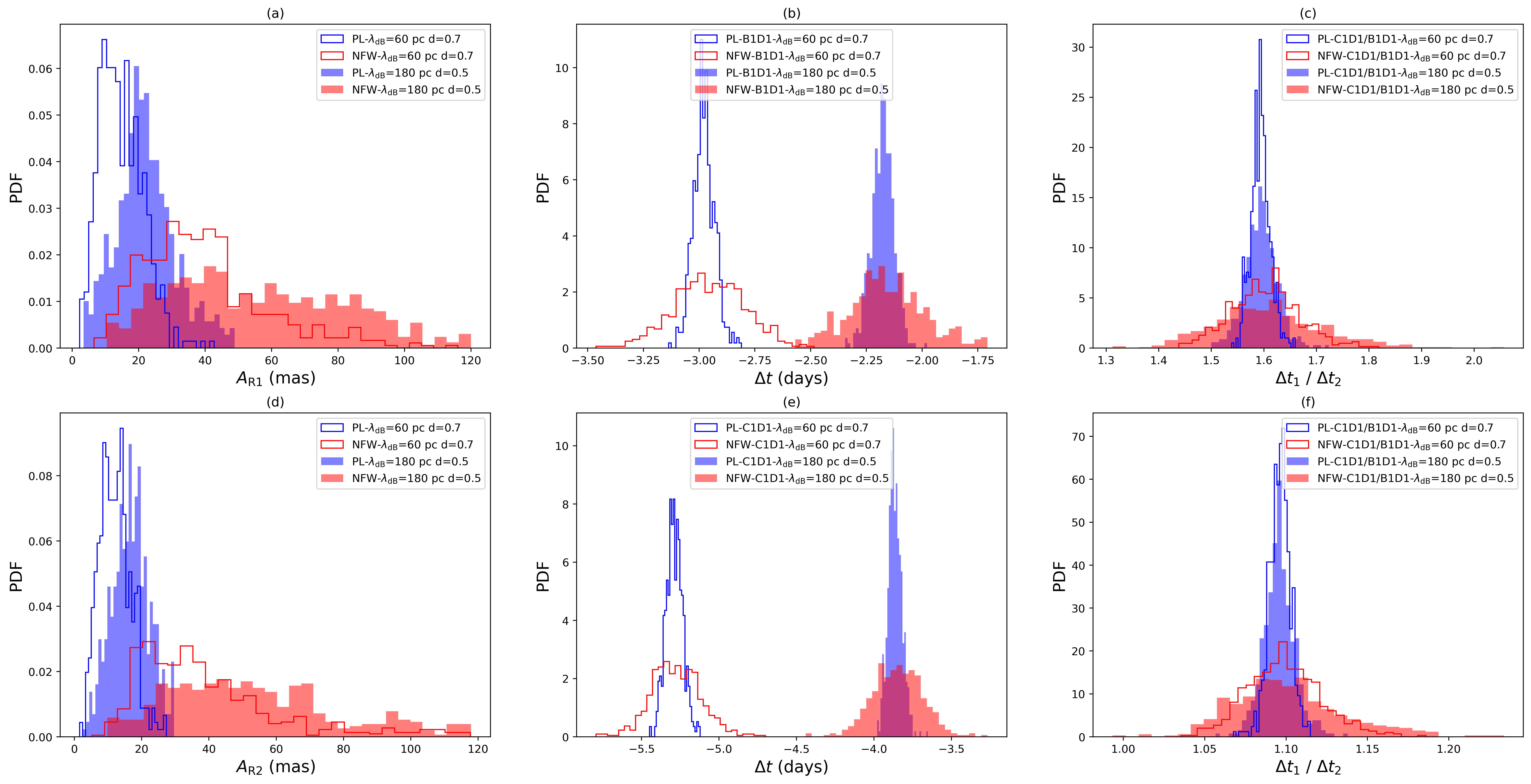}
    \caption{Histograms of the position anomaly $A$ - panels (a,d), time delays (b,e) and time delays ratios (c,f) after adding $\psi \mathrm{DM}$ perturbation to the elliptical PL model and NFW model. Images B1,C1,B2,C2 produced by R1 and R2 jets are considered. 
The position anomaly in this figure is defined in Section \ref{3} and is different from the position anomaly in the Fig. \ref{pl_nfw_error_from_measurement.png}.}
    \label{pl-nfw-compare.png}
\end{figure*}

In order to illustrate the impact of $\psi \mathrm{DM}$ perturbation on the smooth $\rho \mathrm{DM}$ lens model predictions, we introduced the new position anomaly $\hat{A}$  defined as $\hat{A} =\left[\frac{1}{4}\sum_{i=1}^{4}(\mathbf{x}_{i}^{\psi\text{DM}} - \mathbf{x}_{i}^{\varrho\text{DM}} ) ^ { 2 }\right]^{1/2}$. At the same time, we should note that due to the perturbation of caustics by $\psi \mathrm{DM}$, two radio sources at the best fitted location may produce other numbers of images. We ignored this effect and considered only such perturbations that could produce four images. Moreover, we also rejected quads, which manifestly deviate from the observed configuration. Sample size per each combination of $(\lambda_\mathrm{dB}, d)$ parameters tested was about $N=500$ simulations.

\subsection{Time-delay observability}\label{3.3}

In Fig. \ref{change_lambda.png}, we compare distributions of the position anomaly $\hat{A}$ and the position anomaly $A$ used in Section\ref{2.1}. First anomaly captures the effect of $\psi \mathrm{DM}$ perturbations on the predictions of the smooth $\rho \mathrm{DM}$ lens model (elliptical PL profile in this case). The second one is the actual mismatch between the positions predicted by the best fitted $\rho \mathrm{DM}$ model and observed positions of images. Comparison between these two could allow us to determine whether the realistic magnitude of $\psi \mathrm{DM}$ perturbation can explain the position anomalies in HS 0810+2554 system.

We see that while $\psi \mathrm{DM}$ can barely explain the position anomalies of HS 0810+2554, the standard deviation of the time delay anomaly caused by $\psi \mathrm{DM}$ is consistent with the standard deviation of the error caused by the uncertainty of smooth $\rho \mathrm{DM}$ lens model.
The standard deviation of the time delay ratio anomaly caused by $\psi \mathrm{DM}$ is much bigger than the standard deviation of the error caused by the uncertainty of smooth $\rho \mathrm{DM}$ lens model.
We can see that there is a more than 30\% probability of anomalous time delays ratios in Fig.\ref{change_lambda.png} (c). Likewise, in Fig.\ref{change_lambda.png} (f), there is a more than 50\% probability of anomalous time delays ratios.
As we increase the perturbation intensity, the standard deviations of both the time delay anomaly and the time delay ratio anomaly are larger than the standard deviation of the error caused by the standard deviation of the error caused by the uncertainty of smooth $\rho \mathrm{DM}$ lens model.
Compared with time delay anomalies, time delays ratios anomalies are easier to detect.

Considering that the error using common methods such as the light curve of active galactic nuclei to mearsure strong lensing time delays $\sim \mathrm{day}$, but time delays anomalies caused by $\psi \mathrm{DM}$ usually $< 1\;\mathrm{day}$.
This results that the existing time delays data of strong gravitational lensing systems measured using traditional methods are basically unable to meet the needs of studying the time delay perturbation of $\psi \mathrm{DM}$.
Fast radio bursts (FRBs) are bursts in the radio band that last $\sim \mathrm{mas}$. Compared with the usual strong gravitational lensing system time delays $\sim \mathrm{day}$, we can accurately measure the time delays of strong lensed FRBs. 
Although no strongly lensed FRBs have been observed yet, it is expected that about a dozen strongly lensed FRBs will be observed in the next few years \citep{gao2022prospects}.

\subsection{Arrival-time ordering of lensed multiple images}\label{3.2}

\begin{figure}
	\centering
	\begin{minipage}{\linewidth}
		\centering
		\includegraphics[width=\linewidth]{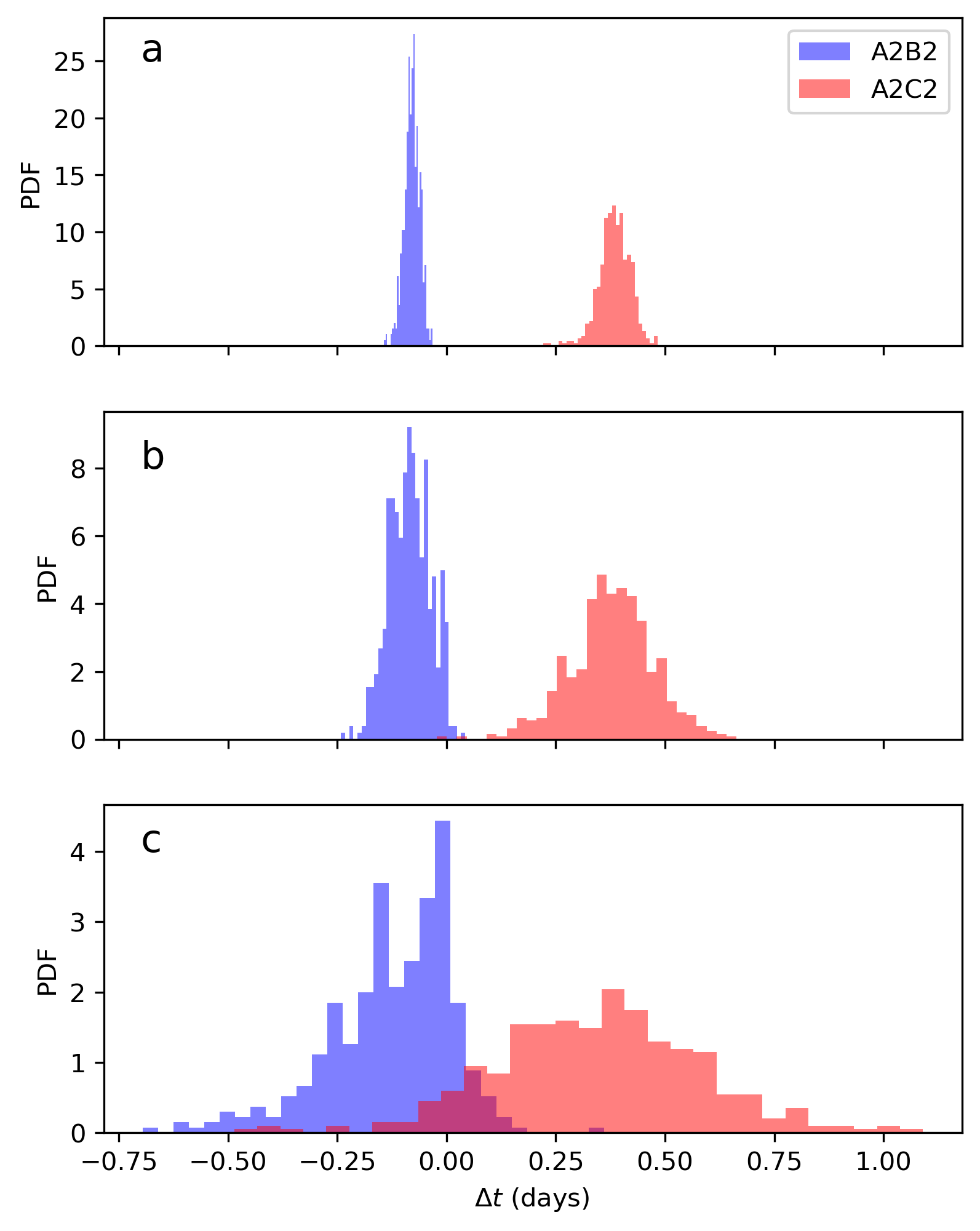}	
	\end{minipage}
    \caption{Time delays anomalies of A2B2 and A2C2. A2 and C2 are minimum points; B2 is the saddle point. Strength of the $\psi$DM perturbation is $d=0.3$ with different de Broglie wavelength:  $\lambda_\mathrm{dB} = 60\;\mathrm{pc}$ panel (a), $\lambda_\mathrm{dB} = 120\;\mathrm{pc}$ panel (b) and $\lambda_\mathrm{dB} = 240\;\mathrm{pc}$ panel (c). }
    \label{Arrival_order_change_60_120_240.png}
\end{figure}

According to strong gravitational lensing theory, strong lensing system with quadruple images would have two saddle points and two minimum points in the time delay surface, of which the minimum point will arrive before the saddle point \citep{saha2003qualitative}. 
This is determined by the topological structure of strong lensing system. Looking back at Fig. \ref{position.png}, A, C are the minimum points and B, D are the saddle points.
The perturbation of the time-delay surface by the dark matter may change the arrival-time order of some images that arrive at relatively close times. It is very easy to change the arrival-time order between two minimum points or two saddle points, which does not violate the general theory. In the $\rho \mathrm{DM}$ scenario, perturbing the smooth $\rho \mathrm{DM}$ lens model by adding subhalos to cause a change in the arrival-time order between saddle point and minima point is unlikely to occur. In order to be effective, a single large-mass subhalo needs to be placed near the image \citep{keeton2009new}. 
However, due to the general wave nature of $\psi \mathrm{DM}$ in space, there is a certain possibility for relatively strong perturbations to destroy the topological structure of the lens system and change the arrival order between the saddle point and the minimum point. 
In Fig. \ref{Arrival_order_change_60_120_240.png}, one can see the following features. The distribution of time delays between two minimum points A2 and C2 is roughly symmetric but broadens as the perturbation intensity increases. On the other hand, distribution of time delays between the minimum (A2) and the saddle (B2) point is becoming skewed with increasing of the perturbation intensity. 
Moreover, from the panel c in Fig. \ref{Arrival_order_change_60_120_240.png} one can conclude, that there is a 15.4$\%$ probability that the topology of the lens system will be destoryed by $\psi \mathrm{DM}$, causing the change in the arrival order between the saddle point (B2) and the minimum point (A2). 

\subsection{Scaling of the time-delay perturbation with $\lambda_\mathrm{dB}$ } \label{3.1}

In the Fig. \ref{pl-nfw-compare.png}, when elliptical PL profile and NFW profile have the same proportional coefficient $d$ and $\lambda_\mathrm{dB}$, the size of the standard deviation of the perturbation caused by $\psi \mathrm{DM}$ to them is also similar, which verifies our idea in the Section \ref{2.2}. At the same time, we can see that the time delays and time delays ratio are very good normal distributions. Next, we verify the change of time delays with proportional coefficient $d$ and $\lambda_\mathrm{dB}$. In the following simulations we all use elliptical PL profile.

\begin{figure}
	\centering
	\begin{minipage}{\linewidth}
		\centering
		\includegraphics[width=\linewidth]{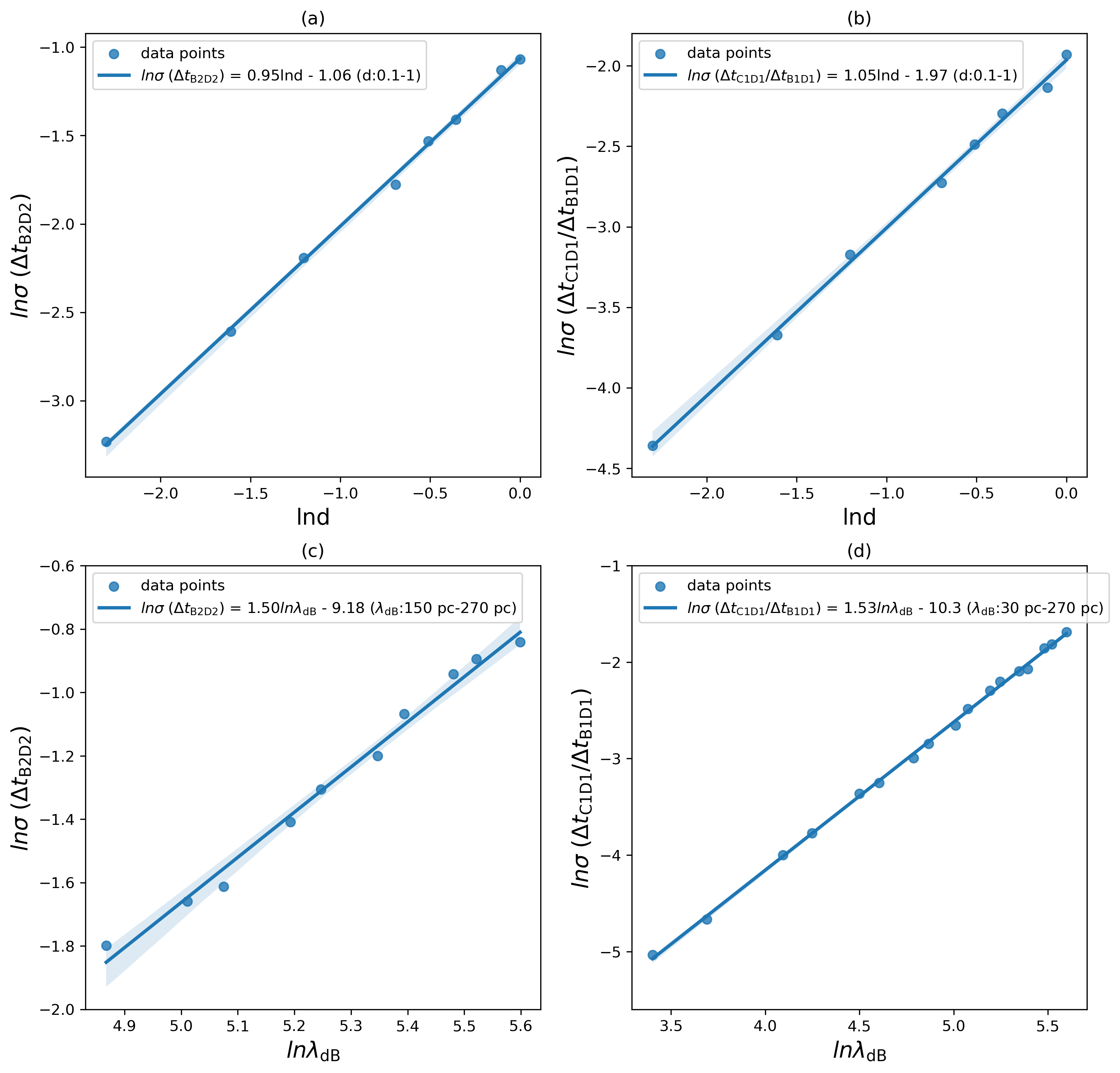}	
	\end{minipage}
    \caption{Plot demonstrating how the magnitude of $\psi \mathrm{DM}$ perturbation changes with parameters $d$ and $\lambda_\mathrm{dB}$. One can see that $\sigma_{\Delta t}\propto d\lambda_\mathrm{dB}^{1.5}$ and $\sigma_{\Delta t_1/\Delta t_2}\propto d\lambda_\mathrm{dB}^{1.5}$ when $\lambda_\mathrm{dB}$ is not too small. In panels (a) and (b) de Broglie wavelength is fixed as $\lambda_\mathrm{dB} = 180\;\mathrm{pc}$; in panels (c) and (d) proportionality coefficient is fixed as $d=0.3$. }
    \label{fit_d_lambda.png}
\end{figure}

\begin{figure}
	\centering
	\begin{minipage}{\linewidth}
		\centering
		\includegraphics[width=\linewidth]{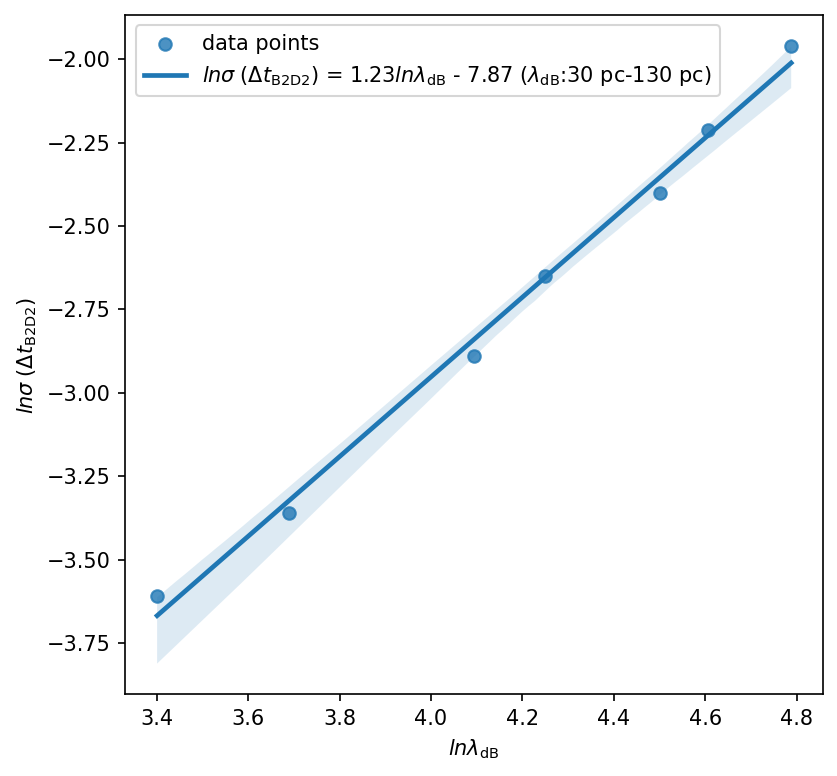}	
	\end{minipage}
    \caption{Plot showing that at smaller $\lambda_\mathrm{dB}$ scales the scaling exponent of $\sigma_{\Delta t}$ is closer to 1 than 1.5. Proportionality coefficient is fixed at $d = 0.3$. }
    \label{fit_d_lambda2.png}
\end{figure}

In Fig. \ref{fit_d_lambda.png} a and b, we can see $\sigma_{\Delta t}\propto d$ and $\sigma_{\Delta t_1/\Delta t_2}\propto d$.
In Fig. \ref{fit_d_lambda.png} d, we can get $\sigma_{\Delta t_1/\Delta t_2}\propto \lambda_\mathrm{dB}^{1.5}$.
In Fig. \ref{fit_d_lambda2.png}, we know that when $\lambda_\mathrm{dB}$ is small, $\sigma_{\Delta t}\propto \lambda_\mathrm{dB}^{1.2}$, the exponent is closer to 1 than 1.5.
As $\lambda_\mathrm{dB}$ increases, the exponent approaches 1.5 (Fig. \ref{fit_d_lambda.png} c). 
This confirms the analytical conclusion about time delay and time delay ratio we obtained in Section \ref{2.3}.

\section{Conclusions and Discussions}\label{4}
Image positions and flux ratios are the most commonly used observational quantities for strong lensing, with which one could distinguish different smooth lens potentials, for example, a power-law model or a NFW model. Time delay measurements are becoming more and more popular thanks to the dedicated follow-up monitoring and the forthcoming large time-domain surveys. Strong lensing of transient events \citep{2019RPPh...82l6901O,2022ChPhL..39k9801L,2024SSRv..220...13S} would be a game changer regarding the precision of strong lensing time delay measurements.  In addition to positions and flux ratios, accurate time delay data can help us distinguish different lens models, for example by breaking the radial profile degeneracy that appears in the relevant lens modeling \citep{keeton2009new,schneider2013source}. 

In this paper, we focused on the perturbation of time delays in strong lensing systems by $\psi \mathrm{DM}$ structures. In order be specific and have an opportunity to relate our assessment to a real case, we studied a particularly well studied strong lensing system HS 0810+2554. After determining a smooth $\rho \mathrm{DM}$ lens model of this system by the MCMC method we quantified the position anomalies left and calculated the expected time delay and time delay ratio anomalies. While the position anomalies could be confronted with observations, taking advantage of very precise measurements of image locations in radio band using VLBI, time delays for this system are currently not available. 

However, we demonstrated clearly that while PL and NFW smooth models are not distinguishable by position anomalies, yet time delay breaks this degeneracy. 
Next, we formulated theoretical expectations regarding time delay perturbations due to 
$\psi \mathrm{DM}$ and tested them in simulated data. Simulations comprised of smooth $\rho$DM models (PL and NFW) perturbed by $\psi$DM, where perturbation was quantified as a standard deviation of a Gaussian random field proportional to the de Broglie wavelength $\lambda_\mathrm{dB}$ and to $\sigma_\rho =d \, \rho_{\mathrm{smooth}}$. Simulations validated our theoretical expectations that $\psi \mathrm{DM}$ induced perturbations follow the following scaling: $\sigma_{\Delta t}\propto d\lambda_\mathrm{dB}^{1.5}$, $\sigma_{\Delta t_1/\Delta t_2}\propto d\lambda_\mathrm{dB}^{1.5}$. We also found that $\psi \mathrm{DM}$ perturbations exceeding certain (realistic) magnitude are able to disrupt the topology of strong lensing system critical curves and caustics, thus changing the arrival-time order of the saddle point and minimum point in the time delay surface.

Time delays of multiply-lensed images are usually measured using light curves of strongly lensed active galactic nuclei (AGN) \citep{2015ApJ...800...11L}, but the the uncertainty of such measurements is usually higher than (at best could be close to) values of time delay anomalies caused by $\psi \mathrm{DM}$. Luckily, strongly lensed transients \citep{2022ChPhL..39k9801L,2018NatCo...9.3833L} including supernovae, gravitational waves, gamma ray bursts and fast radio bursts detected in the time-domain era will provide accurate time delay measurements nearly without uncertainties \citep{liao2017precision} due to their transient nature compared to the typical time delay $\sim$days to $\sim$ years. Thus, there is an opportunity to study time delays anomalies caused by the $\psi \mathrm{DM}$ using these time delay data of strongly lensed transients. One use the time delay anomalies or time delay ratio anomalies to evaluate the parameters in $\psi \mathrm{DM}$. 
More specifically, time delay difference being a signature of the 
change in the arrival-time order of the images of the strong gravitational lensing system (between the saddle point and the minimum point), would be too small to be revealed in strong lensing systems where time delays are measured in a traditional way. 
However, the precise measurement of time delays in strongly lensed transient systems would reveal the effect of changes in arrival-time order. With the subsequent accumulation of time delays data, one would be able to evaluate the parameters of $\psi \mathrm{DM}$ based on the proportion of changes in arrival-time order of the images in the data. 
Furthermore, the methodology we develop in this work can also be appied in studying different dark matter models with substructures \citep{gilman2020warm,dike2023strong,laroche2022quantum}.

\section*{Acknowledgements}
We thank Shuxun Tian for polishing this article.
KL was supported by National Natural Science Foundation of China (NSFC) No. 12222302.
\appendix*

\nocite{*}

\bibliography{apssamp}

\begin{thebibliography}{48}%
\makeatletter
\providecommand \@ifxundefined [1]{%
 \@ifx{#1\undefined}
}%
\providecommand \@ifnum [1]{%
 \ifnum #1\expandafter \@firstoftwo
 \else \expandafter \@secondoftwo
 \fi
}%
\providecommand \@ifx [1]{%
 \ifx #1\expandafter \@firstoftwo
 \else \expandafter \@secondoftwo
 \fi
}%
\providecommand \natexlab [1]{#1}%
\providecommand \enquote  [1]{``#1''}%
\providecommand \bibnamefont  [1]{#1}%
\providecommand \bibfnamefont [1]{#1}%
\providecommand \citenamefont [1]{#1}%
\providecommand \href@noop [0]{\@secondoftwo}%
\providecommand \href [0]{\begingroup \@sanitize@url \@href}%
\providecommand \@href[1]{\@@startlink{#1}\@@href}%
\providecommand \@@href[1]{\endgroup#1\@@endlink}%
\providecommand \@sanitize@url [0]{\catcode `\\12\catcode `\$12\catcode `\&12\catcode `\#12\catcode `\^12\catcode `\_12\catcode `\%12\relax}%
\providecommand \@@startlink[1]{}%
\providecommand \@@endlink[0]{}%
\providecommand \url  [0]{\begingroup\@sanitize@url \@url }%
\providecommand \@url [1]{\endgroup\@href {#1}{\urlprefix }}%
\providecommand \urlprefix  [0]{URL }%
\providecommand \Eprint [0]{\href }%
\providecommand \doibase [0]{https://doi.org/}%
\providecommand \selectlanguage [0]{\@gobble}%
\providecommand \bibinfo  [0]{\@secondoftwo}%
\providecommand \bibfield  [0]{\@secondoftwo}%
\providecommand \translation [1]{[#1]}%
\providecommand \BibitemOpen [0]{}%
\providecommand \bibitemStop [0]{}%
\providecommand \bibitemNoStop [0]{.\EOS\space}%
\providecommand \EOS [0]{\spacefactor3000\relax}%
\providecommand \BibitemShut  [1]{\csname bibitem#1\endcsname}%
\let\auto@bib@innerbib\@empty
\bibitem [{\citenamefont {{Springel}}\ \emph {et~al.}(2018)\citenamefont {{Springel}}, \citenamefont {{Pakmor}}, \citenamefont {{Pillepich}}, \citenamefont {{Weinberger}}, \citenamefont {{Nelson}}, \citenamefont {{Hernquist}}, \citenamefont {{Vogelsberger}}, \citenamefont {{Genel}}, \citenamefont {{Torrey}}, \citenamefont {{Marinacci}},\ and\ \citenamefont {{Naiman}}}]{10.1093/mnras/stx3304}%
  \BibitemOpen
  \bibfield  {author} {\bibinfo {author} {\bibfnamefont {V.}~\bibnamefont {{Springel}}}, \bibinfo {author} {\bibfnamefont {R.}~\bibnamefont {{Pakmor}}}, \bibinfo {author} {\bibfnamefont {A.}~\bibnamefont {{Pillepich}}}, \bibinfo {author} {\bibfnamefont {R.}~\bibnamefont {{Weinberger}}}, \bibinfo {author} {\bibfnamefont {D.}~\bibnamefont {{Nelson}}}, \bibinfo {author} {\bibfnamefont {L.}~\bibnamefont {{Hernquist}}}, \bibinfo {author} {\bibfnamefont {M.}~\bibnamefont {{Vogelsberger}}}, \bibinfo {author} {\bibfnamefont {S.}~\bibnamefont {{Genel}}}, \bibinfo {author} {\bibfnamefont {P.}~\bibnamefont {{Torrey}}}, \bibinfo {author} {\bibfnamefont {F.}~\bibnamefont {{Marinacci}}},\ and\ \bibinfo {author} {\bibfnamefont {J.}~\bibnamefont {{Naiman}}},\ }\bibfield  {title} {\bibinfo {title} {{First results from the IllustrisTNG simulations: matter and galaxy clustering}},\ }\href {https://doi.org/10.1093/mnras/stx3304} {\bibfield  {journal} {\bibinfo  {journal} {Monthly Notices of the Royal Astronomical Society}\
  }\textbf {\bibinfo {volume} {475}},\ \bibinfo {pages} {676} (\bibinfo {year} {2018})},\ \Eprint {https://arxiv.org/abs/1707.03397} {arXiv:1707.03397 [astro-ph.GA]} \BibitemShut {NoStop}%
\bibitem [{\citenamefont {{Bullock}}\ and\ \citenamefont {{Boylan-Kolchin}}(2017)}]{bullock2017small}%
  \BibitemOpen
  \bibfield  {author} {\bibinfo {author} {\bibfnamefont {J.~S.}\ \bibnamefont {{Bullock}}}\ and\ \bibinfo {author} {\bibfnamefont {M.}~\bibnamefont {{Boylan-Kolchin}}},\ }\bibfield  {title} {\bibinfo {title} {{Small-Scale Challenges to the {\ensuremath{\Lambda}}CDM Paradigm}},\ }\href {https://doi.org/10.1146/annurev-astro-091916-055313} {\bibfield  {journal} {\bibinfo  {journal} {Annual Review of Astronomy and Astrophysics}\ }\textbf {\bibinfo {volume} {55}},\ \bibinfo {pages} {343} (\bibinfo {year} {2017})},\ \Eprint {https://arxiv.org/abs/1707.04256} {arXiv:1707.04256 [astro-ph.CO]} \BibitemShut {NoStop}%
\bibitem [{\citenamefont {{Chan}}\ \emph {et~al.}(2015)\citenamefont {{Chan}}, \citenamefont {{Kere{\v{s}}}}, \citenamefont {{O{\~n}orbe}}, \citenamefont {{Hopkins}}, \citenamefont {{Muratov}}, \citenamefont {{Faucher-Gigu{\`e}re}},\ and\ \citenamefont {{Quataert}}}]{chan2015impact}%
  \BibitemOpen
  \bibfield  {author} {\bibinfo {author} {\bibfnamefont {T.~K.}\ \bibnamefont {{Chan}}}, \bibinfo {author} {\bibfnamefont {D.}~\bibnamefont {{Kere{\v{s}}}}}, \bibinfo {author} {\bibfnamefont {J.}~\bibnamefont {{O{\~n}orbe}}}, \bibinfo {author} {\bibfnamefont {P.~F.}\ \bibnamefont {{Hopkins}}}, \bibinfo {author} {\bibfnamefont {A.~L.}\ \bibnamefont {{Muratov}}}, \bibinfo {author} {\bibfnamefont {C.~A.}\ \bibnamefont {{Faucher-Gigu{\`e}re}}},\ and\ \bibinfo {author} {\bibfnamefont {E.}~\bibnamefont {{Quataert}}},\ }\bibfield  {title} {\bibinfo {title} {{The impact of baryonic physics on the structure of dark matter haloes: the view from the FIRE cosmological simulations}},\ }\href {https://doi.org/10.1093/mnras/stv2165} {\bibfield  {journal} {\bibinfo  {journal} {Monthly Notices of the Royal Astronomical Society}\ }\textbf {\bibinfo {volume} {454}},\ \bibinfo {pages} {2981} (\bibinfo {year} {2015})},\ \Eprint {https://arxiv.org/abs/1507.02282} {arXiv:1507.02282 [astro-ph.GA]} \BibitemShut {NoStop}%
\bibitem [{\citenamefont {{Jungman}}\ \emph {et~al.}(1996)\citenamefont {{Jungman}}, \citenamefont {{Kamionkowski}},\ and\ \citenamefont {{Griest}}}]{jungman1996supersymmetric}%
  \BibitemOpen
  \bibfield  {author} {\bibinfo {author} {\bibfnamefont {G.}~\bibnamefont {{Jungman}}}, \bibinfo {author} {\bibfnamefont {M.}~\bibnamefont {{Kamionkowski}}},\ and\ \bibinfo {author} {\bibfnamefont {K.}~\bibnamefont {{Griest}}},\ }\bibfield  {title} {\bibinfo {title} {{Supersymmetric dark matter}},\ }\href {https://doi.org/10.1016/0370-1573(95)00058-5} {\bibfield  {journal} {\bibinfo  {journal} {Physics Reports}\ }\textbf {\bibinfo {volume} {267}},\ \bibinfo {pages} {195} (\bibinfo {year} {1996})},\ \Eprint {https://arxiv.org/abs/hep-ph/9506380} {arXiv:hep-ph/9506380 [hep-ph]} \BibitemShut {NoStop}%
\bibitem [{\citenamefont {{Arvanitaki}}\ \emph {et~al.}(2010)\citenamefont {{Arvanitaki}}, \citenamefont {{Dimopoulos}}, \citenamefont {{Dubovsky}}, \citenamefont {{Kaloper}},\ and\ \citenamefont {{March-Russell}}}]{arvanitaki2010string}%
  \BibitemOpen
  \bibfield  {author} {\bibinfo {author} {\bibfnamefont {A.}~\bibnamefont {{Arvanitaki}}}, \bibinfo {author} {\bibfnamefont {S.}~\bibnamefont {{Dimopoulos}}}, \bibinfo {author} {\bibfnamefont {S.}~\bibnamefont {{Dubovsky}}}, \bibinfo {author} {\bibfnamefont {N.}~\bibnamefont {{Kaloper}}},\ and\ \bibinfo {author} {\bibfnamefont {J.}~\bibnamefont {{March-Russell}}},\ }\bibfield  {title} {\bibinfo {title} {{String axiverse}},\ }\href {https://doi.org/10.1103/PhysRevD.81.123530} {\bibfield  {journal} {\bibinfo  {journal} {Physical Review D}\ }\textbf {\bibinfo {volume} {81}},\ \bibinfo {eid} {123530} (\bibinfo {year} {2010})},\ \Eprint {https://arxiv.org/abs/0905.4720} {arXiv:0905.4720 [hep-th]} \BibitemShut {NoStop}%
\bibitem [{\citenamefont {{Svrcek}}\ and\ \citenamefont {{Witten}}(2006)}]{svrcek2006axions}%
  \BibitemOpen
  \bibfield  {author} {\bibinfo {author} {\bibfnamefont {P.}~\bibnamefont {{Svrcek}}}\ and\ \bibinfo {author} {\bibfnamefont {E.}~\bibnamefont {{Witten}}},\ }\bibfield  {title} {\bibinfo {title} {{Axions in string theory}},\ }\href {https://doi.org/10.1088/1126-6708/2006/06/051} {\bibfield  {journal} {\bibinfo  {journal} {Journal of High Energy Physics}\ }\textbf {\bibinfo {volume} {2006}},\ \bibinfo {eid} {051} (\bibinfo {year} {2006})},\ \Eprint {https://arxiv.org/abs/hep-th/0605206} {arXiv:hep-th/0605206 [hep-th]} \BibitemShut {NoStop}%
\bibitem [{\citenamefont {{Marsh}}(2016)}]{marsh2016axion}%
  \BibitemOpen
  \bibfield  {author} {\bibinfo {author} {\bibfnamefont {D.~J.~E.}\ \bibnamefont {{Marsh}}},\ }\bibfield  {title} {\bibinfo {title} {{Axion cosmology}},\ }\href {https://doi.org/10.1016/j.physrep.2016.06.005} {\bibfield  {journal} {\bibinfo  {journal} {Physics Reports}\ }\textbf {\bibinfo {volume} {643}},\ \bibinfo {pages} {1} (\bibinfo {year} {2016})},\ \Eprint {https://arxiv.org/abs/1510.07633} {arXiv:1510.07633 [astro-ph.CO]} \BibitemShut {NoStop}%
\bibitem [{\citenamefont {{Hu}}\ \emph {et~al.}(2000)\citenamefont {{Hu}}, \citenamefont {{Barkana}},\ and\ \citenamefont {{Gruzinov}}}]{hu2000fuzzy}%
  \BibitemOpen
  \bibfield  {author} {\bibinfo {author} {\bibfnamefont {W.}~\bibnamefont {{Hu}}}, \bibinfo {author} {\bibfnamefont {R.}~\bibnamefont {{Barkana}}},\ and\ \bibinfo {author} {\bibfnamefont {A.}~\bibnamefont {{Gruzinov}}},\ }\bibfield  {title} {\bibinfo {title} {{Fuzzy Cold Dark Matter: The Wave Properties of Ultralight Particles}},\ }\href {https://doi.org/10.1103/PhysRevLett.85.1158} {\bibfield  {journal} {\bibinfo  {journal} {Physical Review Letters}\ }\textbf {\bibinfo {volume} {85}},\ \bibinfo {pages} {1158} (\bibinfo {year} {2000})},\ \Eprint {https://arxiv.org/abs/astro-ph/0003365} {arXiv:astro-ph/0003365 [astro-ph]} \BibitemShut {NoStop}%
\bibitem [{\citenamefont {{Schive}}\ \emph {et~al.}(2014{\natexlab{a}})\citenamefont {{Schive}}, \citenamefont {{Chiueh}},\ and\ \citenamefont {{Broadhurst}}}]{schive2014cosmic}%
  \BibitemOpen
  \bibfield  {author} {\bibinfo {author} {\bibfnamefont {H.-Y.}\ \bibnamefont {{Schive}}}, \bibinfo {author} {\bibfnamefont {T.}~\bibnamefont {{Chiueh}}},\ and\ \bibinfo {author} {\bibfnamefont {T.}~\bibnamefont {{Broadhurst}}},\ }\bibfield  {title} {\bibinfo {title} {{Cosmic structure as the quantum interference of a coherent dark wave}},\ }\href {https://doi.org/10.1038/nphys2996} {\bibfield  {journal} {\bibinfo  {journal} {Nature Physics}\ }\textbf {\bibinfo {volume} {10}},\ \bibinfo {pages} {496} (\bibinfo {year} {2014}{\natexlab{a}})},\ \Eprint {https://arxiv.org/abs/1406.6586} {arXiv:1406.6586 [astro-ph.GA]} \BibitemShut {NoStop}%
\bibitem [{\citenamefont {{Schive}}\ \emph {et~al.}(2014{\natexlab{b}})\citenamefont {{Schive}}, \citenamefont {{Liao}}, \citenamefont {{Woo}}, \citenamefont {{Wong}}, \citenamefont {{Chiueh}}, \citenamefont {{Broadhurst}},\ and\ \citenamefont {{Hwang}}}]{schive2014understanding}%
  \BibitemOpen
  \bibfield  {author} {\bibinfo {author} {\bibfnamefont {H.-Y.}\ \bibnamefont {{Schive}}}, \bibinfo {author} {\bibfnamefont {M.-H.}\ \bibnamefont {{Liao}}}, \bibinfo {author} {\bibfnamefont {T.-P.}\ \bibnamefont {{Woo}}}, \bibinfo {author} {\bibfnamefont {S.-K.}\ \bibnamefont {{Wong}}}, \bibinfo {author} {\bibfnamefont {T.}~\bibnamefont {{Chiueh}}}, \bibinfo {author} {\bibfnamefont {T.}~\bibnamefont {{Broadhurst}}},\ and\ \bibinfo {author} {\bibfnamefont {W.~Y.~P.}\ \bibnamefont {{Hwang}}},\ }\bibfield  {title} {\bibinfo {title} {{Understanding the Core-Halo Relation of Quantum Wave Dark Matter from 3D Simulations}},\ }\href {https://doi.org/10.1103/PhysRevLett.113.261302} {\bibfield  {journal} {\bibinfo  {journal} {Physical review letters}\ }\textbf {\bibinfo {volume} {113}},\ \bibinfo {eid} {261302} (\bibinfo {year} {2014}{\natexlab{b}})},\ \Eprint {https://arxiv.org/abs/1407.7762} {arXiv:1407.7762 [astro-ph.GA]} \BibitemShut {NoStop}%
\bibitem [{\citenamefont {{Hui}}(2021)}]{hui2021wave}%
  \BibitemOpen
  \bibfield  {author} {\bibinfo {author} {\bibfnamefont {L.}~\bibnamefont {{Hui}}},\ }\bibfield  {title} {\bibinfo {title} {{Wave Dark Matter}},\ }\href {https://doi.org/10.1146/annurev-astro-120920-010024} {\bibfield  {journal} {\bibinfo  {journal} {Annual Review of Astronomy and Astrophysics}\ }\textbf {\bibinfo {volume} {59}},\ \bibinfo {pages} {247} (\bibinfo {year} {2021})},\ \Eprint {https://arxiv.org/abs/2101.11735} {arXiv:2101.11735 [astro-ph.CO]} \BibitemShut {NoStop}%
\bibitem [{\citenamefont {{Mocz}}\ \emph {et~al.}(2019)\citenamefont {{Mocz}}, \citenamefont {{Fialkov}}, \citenamefont {{Vogelsberger}}, \citenamefont {{Becerra}}, \citenamefont {{Amin}}, \citenamefont {{Bose}}, \citenamefont {{Boylan-Kolchin}}, \citenamefont {{Chavanis}}, \citenamefont {{Hernquist}}, \citenamefont {{Lancaster}}, \citenamefont {{Marinacci}}, \citenamefont {{Robles}},\ and\ \citenamefont {{Zavala}}}]{mocz2019first}%
  \BibitemOpen
  \bibfield  {author} {\bibinfo {author} {\bibfnamefont {P.}~\bibnamefont {{Mocz}}}, \bibinfo {author} {\bibfnamefont {A.}~\bibnamefont {{Fialkov}}}, \bibinfo {author} {\bibfnamefont {M.}~\bibnamefont {{Vogelsberger}}}, \bibinfo {author} {\bibfnamefont {F.}~\bibnamefont {{Becerra}}}, \bibinfo {author} {\bibfnamefont {M.~A.}\ \bibnamefont {{Amin}}}, \bibinfo {author} {\bibfnamefont {S.}~\bibnamefont {{Bose}}}, \bibinfo {author} {\bibfnamefont {M.}~\bibnamefont {{Boylan-Kolchin}}}, \bibinfo {author} {\bibfnamefont {P.-H.}\ \bibnamefont {{Chavanis}}}, \bibinfo {author} {\bibfnamefont {L.}~\bibnamefont {{Hernquist}}}, \bibinfo {author} {\bibfnamefont {L.}~\bibnamefont {{Lancaster}}}, \bibinfo {author} {\bibfnamefont {F.}~\bibnamefont {{Marinacci}}}, \bibinfo {author} {\bibfnamefont {V.~H.}\ \bibnamefont {{Robles}}},\ and\ \bibinfo {author} {\bibfnamefont {J.}~\bibnamefont {{Zavala}}},\ }\bibfield  {title} {\bibinfo {title} {{First Star-Forming Structures in Fuzzy Cosmic Filaments}},\ }\href
  {https://doi.org/10.1103/PhysRevLett.123.141301} {\bibfield  {journal} {\bibinfo  {journal} {Physical review letters}\ }\textbf {\bibinfo {volume} {123}},\ \bibinfo {eid} {141301} (\bibinfo {year} {2019})},\ \Eprint {https://arxiv.org/abs/1910.01653} {arXiv:1910.01653 [astro-ph.GA]} \BibitemShut {NoStop}%
\bibitem [{\citenamefont {{Woo}}\ and\ \citenamefont {{Chiueh}}(2009)}]{woo2009high}%
  \BibitemOpen
  \bibfield  {author} {\bibinfo {author} {\bibfnamefont {T.-P.}\ \bibnamefont {{Woo}}}\ and\ \bibinfo {author} {\bibfnamefont {T.}~\bibnamefont {{Chiueh}}},\ }\bibfield  {title} {\bibinfo {title} {{High-Resolution Simulation on Structure Formation with Extremely Light Bosonic Dark Matter}},\ }\href {https://doi.org/10.1088/0004-637X/697/1/850} {\bibfield  {journal} {\bibinfo  {journal} {The Astrophysical Journal}\ }\textbf {\bibinfo {volume} {697}},\ \bibinfo {pages} {850} (\bibinfo {year} {2009})},\ \Eprint {https://arxiv.org/abs/0806.0232} {arXiv:0806.0232 [astro-ph]} \BibitemShut {NoStop}%
\bibitem [{\citenamefont {{Dalal}}\ and\ \citenamefont {{Kravtsov}}(2022)}]{dalal2022excluding}%
  \BibitemOpen
  \bibfield  {author} {\bibinfo {author} {\bibfnamefont {N.}~\bibnamefont {{Dalal}}}\ and\ \bibinfo {author} {\bibfnamefont {A.}~\bibnamefont {{Kravtsov}}},\ }\bibfield  {title} {\bibinfo {title} {{Excluding fuzzy dark matter with sizes and stellar kinematics of ultrafaint dwarf galaxies}},\ }\href {https://doi.org/10.1103/PhysRevD.106.063517} {\bibfield  {journal} {\bibinfo  {journal} {Physical Review D}\ }\textbf {\bibinfo {volume} {106}},\ \bibinfo {eid} {063517} (\bibinfo {year} {2022})}\BibitemShut {NoStop}%
\bibitem [{\citenamefont {{Nadler}}\ \emph {et~al.}(2021)\citenamefont {{Nadler}}, \citenamefont {{Drlica-Wagner}}, \citenamefont {{Bechtol}},\ and\ \citenamefont {{et al.}}}]{nadler2021constraints}%
  \BibitemOpen
  \bibfield  {author} {\bibinfo {author} {\bibfnamefont {E.~O.}\ \bibnamefont {{Nadler}}}, \bibinfo {author} {\bibfnamefont {A.}~\bibnamefont {{Drlica-Wagner}}}, \bibinfo {author} {\bibfnamefont {K.}~\bibnamefont {{Bechtol}}},\ and\ \bibinfo {author} {\bibnamefont {{et al.}}},\ }\bibfield  {title} {\bibinfo {title} {{Constraints on Dark Matter Properties from Observations of Milky Way Satellite Galaxies}},\ }\href {https://doi.org/10.1103/PhysRevLett.126.091101} {\bibfield  {journal} {\bibinfo  {journal} {Physical review letters}\ }\textbf {\bibinfo {volume} {126}},\ \bibinfo {eid} {091101} (\bibinfo {year} {2021})},\ \Eprint {https://arxiv.org/abs/2008.00022} {arXiv:2008.00022 [astro-ph.CO]} \BibitemShut {NoStop}%
\bibitem [{\citenamefont {{Schutz}}(2020)}]{schutz2020subhalo}%
  \BibitemOpen
  \bibfield  {author} {\bibinfo {author} {\bibfnamefont {K.}~\bibnamefont {{Schutz}}},\ }\bibfield  {title} {\bibinfo {title} {{Subhalo mass function and ultralight bosonic dark matter}},\ }\href {https://doi.org/10.1103/PhysRevD.101.123026} {\bibfield  {journal} {\bibinfo  {journal} {Physical Review D}\ }\textbf {\bibinfo {volume} {101}},\ \bibinfo {eid} {123026} (\bibinfo {year} {2020})},\ \Eprint {https://arxiv.org/abs/2001.05503} {arXiv:2001.05503 [astro-ph.CO]} \BibitemShut {NoStop}%
\bibitem [{\citenamefont {{Amruth}}\ \emph {et~al.}(2023)\citenamefont {{Amruth}}, \citenamefont {{Broadhurst}}, \citenamefont {{Lim}}, \citenamefont {{Oguri}}, \citenamefont {{Smoot}}, \citenamefont {{Diego}}, \citenamefont {{Leung}}, \citenamefont {{Emami}}, \citenamefont {{Li}}, \citenamefont {{Chiueh}}, \citenamefont {{Schive}}, \citenamefont {{Yeung}},\ and\ \citenamefont {{Li}}}]{amruth2023einstein}%
  \BibitemOpen
  \bibfield  {author} {\bibinfo {author} {\bibfnamefont {A.}~\bibnamefont {{Amruth}}}, \bibinfo {author} {\bibfnamefont {T.}~\bibnamefont {{Broadhurst}}}, \bibinfo {author} {\bibfnamefont {J.}~\bibnamefont {{Lim}}}, \bibinfo {author} {\bibfnamefont {M.}~\bibnamefont {{Oguri}}}, \bibinfo {author} {\bibfnamefont {G.~F.}\ \bibnamefont {{Smoot}}}, \bibinfo {author} {\bibfnamefont {J.~M.}\ \bibnamefont {{Diego}}}, \bibinfo {author} {\bibfnamefont {E.}~\bibnamefont {{Leung}}}, \bibinfo {author} {\bibfnamefont {R.}~\bibnamefont {{Emami}}}, \bibinfo {author} {\bibfnamefont {J.}~\bibnamefont {{Li}}}, \bibinfo {author} {\bibfnamefont {T.}~\bibnamefont {{Chiueh}}}, \bibinfo {author} {\bibfnamefont {H.-Y.}\ \bibnamefont {{Schive}}}, \bibinfo {author} {\bibfnamefont {M.~C.~H.}\ \bibnamefont {{Yeung}}},\ and\ \bibinfo {author} {\bibfnamefont {S.~K.}\ \bibnamefont {{Li}}},\ }\bibfield  {title} {\bibinfo {title} {{Einstein rings modulated by wavelike dark matter from anomalies in gravitationally lensed images}},\ }\href
  {https://doi.org/10.1038/s41550-023-01943-9} {\bibfield  {journal} {\bibinfo  {journal} {Nature Astronomy}\ }\textbf {\bibinfo {volume} {7}},\ \bibinfo {pages} {736} (\bibinfo {year} {2023})},\ \Eprint {https://arxiv.org/abs/2304.09895} {arXiv:2304.09895 [astro-ph.CO]} \BibitemShut {NoStop}%
\bibitem [{\citenamefont {{Nierenberg}}\ \emph {et~al.}(2020)\citenamefont {{Nierenberg}}, \citenamefont {{Gilman}}, \citenamefont {{Treu}}, \citenamefont {{Brammer}}, \citenamefont {{Birrer}}, \citenamefont {{Moustakas}}, \citenamefont {{Agnello}}, \citenamefont {{Anguita}}, \citenamefont {{Fassnacht}}, \citenamefont {{Motta}}, \citenamefont {{Peter}},\ and\ \citenamefont {{Sluse}}}]{nierenberg2020double}%
  \BibitemOpen
  \bibfield  {author} {\bibinfo {author} {\bibfnamefont {A.~M.}\ \bibnamefont {{Nierenberg}}}, \bibinfo {author} {\bibfnamefont {D.}~\bibnamefont {{Gilman}}}, \bibinfo {author} {\bibfnamefont {T.}~\bibnamefont {{Treu}}}, \bibinfo {author} {\bibfnamefont {G.}~\bibnamefont {{Brammer}}}, \bibinfo {author} {\bibfnamefont {S.}~\bibnamefont {{Birrer}}}, \bibinfo {author} {\bibfnamefont {L.}~\bibnamefont {{Moustakas}}}, \bibinfo {author} {\bibfnamefont {A.}~\bibnamefont {{Agnello}}}, \bibinfo {author} {\bibfnamefont {T.}~\bibnamefont {{Anguita}}}, \bibinfo {author} {\bibfnamefont {C.~D.}\ \bibnamefont {{Fassnacht}}}, \bibinfo {author} {\bibfnamefont {V.}~\bibnamefont {{Motta}}}, \bibinfo {author} {\bibfnamefont {A.~H.~G.}\ \bibnamefont {{Peter}}},\ and\ \bibinfo {author} {\bibfnamefont {D.}~\bibnamefont {{Sluse}}},\ }\bibfield  {title} {\bibinfo {title} {{Double dark matter vision: twice the number of compact-source lenses with narrow-line lensing and the WFC3 grism}},\ }\href {https://doi.org/10.1093/mnras/stz3588}
  {\bibfield  {journal} {\bibinfo  {journal} {Monthly Notices of the Royal Astronomical Society}\ }\textbf {\bibinfo {volume} {492}},\ \bibinfo {pages} {5314} (\bibinfo {year} {2020})},\ \Eprint {https://arxiv.org/abs/1908.06344} {arXiv:1908.06344 [astro-ph.GA]} \BibitemShut {NoStop}%
\bibitem [{\citenamefont {{Keeton}}\ \emph {et~al.}(2003)\citenamefont {{Keeton}}, \citenamefont {{Gaudi}},\ and\ \citenamefont {{Petters}}}]{keeton2003identifying}%
  \BibitemOpen
  \bibfield  {author} {\bibinfo {author} {\bibfnamefont {C.~R.}\ \bibnamefont {{Keeton}}}, \bibinfo {author} {\bibfnamefont {B.~S.}\ \bibnamefont {{Gaudi}}},\ and\ \bibinfo {author} {\bibfnamefont {A.~O.}\ \bibnamefont {{Petters}}},\ }\bibfield  {title} {\bibinfo {title} {{Identifying Lenses with Small-Scale Structure. I. Cusp Lenses}},\ }\href {https://doi.org/10.1086/378934} {\bibfield  {journal} {\bibinfo  {journal} {The Astrophysical Journal}\ }\textbf {\bibinfo {volume} {598}},\ \bibinfo {pages} {138} (\bibinfo {year} {2003})},\ \Eprint {https://arxiv.org/abs/astro-ph/0210318} {arXiv:astro-ph/0210318 [astro-ph]} \BibitemShut {NoStop}%
\bibitem [{\citenamefont {{Kochanek}}\ and\ \citenamefont {{Dalal}}(2004)}]{kochanek2004tests}%
  \BibitemOpen
  \bibfield  {author} {\bibinfo {author} {\bibfnamefont {C.~S.}\ \bibnamefont {{Kochanek}}}\ and\ \bibinfo {author} {\bibfnamefont {N.}~\bibnamefont {{Dalal}}},\ }\bibfield  {title} {\bibinfo {title} {{Tests for Substructure in Gravitational Lenses}},\ }\href {https://doi.org/10.1086/421436} {\bibfield  {journal} {\bibinfo  {journal} {The Astrophysical Journal}\ }\textbf {\bibinfo {volume} {610}},\ \bibinfo {pages} {69} (\bibinfo {year} {2004})},\ \Eprint {https://arxiv.org/abs/astro-ph/0302036} {arXiv:astro-ph/0302036 [astro-ph]} \BibitemShut {NoStop}%
\bibitem [{\citenamefont {{Goldberg}}\ \emph {et~al.}(2010)\citenamefont {{Goldberg}}, \citenamefont {{Chessey}}, \citenamefont {{Harris}},\ and\ \citenamefont {{Richards}}}]{goldberg2010fold}%
  \BibitemOpen
  \bibfield  {author} {\bibinfo {author} {\bibfnamefont {D.~M.}\ \bibnamefont {{Goldberg}}}, \bibinfo {author} {\bibfnamefont {M.~K.}\ \bibnamefont {{Chessey}}}, \bibinfo {author} {\bibfnamefont {W.~B.}\ \bibnamefont {{Harris}}},\ and\ \bibinfo {author} {\bibfnamefont {G.~T.}\ \bibnamefont {{Richards}}},\ }\bibfield  {title} {\bibinfo {title} {{Fold Lens Flux Anomalies: A Geometric Approach}},\ }\href {https://doi.org/10.1088/0004-637X/715/2/793} {\bibfield  {journal} {\bibinfo  {journal} {The Astrophysical Journal}\ }\textbf {\bibinfo {volume} {715}},\ \bibinfo {pages} {793} (\bibinfo {year} {2010})},\ \Eprint {https://arxiv.org/abs/0912.0916} {arXiv:0912.0916 [astro-ph.CO]} \BibitemShut {NoStop}%
\bibitem [{\citenamefont {{Shajib}}\ \emph {et~al.}(2019)\citenamefont {{Shajib}}, \citenamefont {{Birrer}}, \citenamefont {{Treu}},\ and\ \citenamefont {{et al.}}}]{shajib2019every}%
  \BibitemOpen
  \bibfield  {author} {\bibinfo {author} {\bibfnamefont {A.~J.}\ \bibnamefont {{Shajib}}}, \bibinfo {author} {\bibfnamefont {S.}~\bibnamefont {{Birrer}}}, \bibinfo {author} {\bibfnamefont {T.}~\bibnamefont {{Treu}}},\ and\ \bibinfo {author} {\bibnamefont {{et al.}}},\ }\bibfield  {title} {\bibinfo {title} {{Is every strong lens model unhappy in its own way? Uniform modelling of a sample of 13 quadruply+ imaged quasars}},\ }\href {https://doi.org/10.1093/mnras/sty3397} {\bibfield  {journal} {\bibinfo  {journal} {Monthly Notices of the Royal Astronomical Society}\ }\textbf {\bibinfo {volume} {483}},\ \bibinfo {pages} {5649} (\bibinfo {year} {2019})},\ \Eprint {https://arxiv.org/abs/1807.09278} {arXiv:1807.09278 [astro-ph.GA]} \BibitemShut {NoStop}%
\bibitem [{\citenamefont {{Xu}}\ \emph {et~al.}(2015)\citenamefont {{Xu}}, \citenamefont {{Sluse}}, \citenamefont {{Gao}}, \citenamefont {{Wang}}, \citenamefont {{Frenk}}, \citenamefont {{Mao}}, \citenamefont {{Schneider}},\ and\ \citenamefont {{Springel}}}]{xu2015well}%
  \BibitemOpen
  \bibfield  {author} {\bibinfo {author} {\bibfnamefont {D.}~\bibnamefont {{Xu}}}, \bibinfo {author} {\bibfnamefont {D.}~\bibnamefont {{Sluse}}}, \bibinfo {author} {\bibfnamefont {L.}~\bibnamefont {{Gao}}}, \bibinfo {author} {\bibfnamefont {J.}~\bibnamefont {{Wang}}}, \bibinfo {author} {\bibfnamefont {C.}~\bibnamefont {{Frenk}}}, \bibinfo {author} {\bibfnamefont {S.}~\bibnamefont {{Mao}}}, \bibinfo {author} {\bibfnamefont {P.}~\bibnamefont {{Schneider}}},\ and\ \bibinfo {author} {\bibfnamefont {V.}~\bibnamefont {{Springel}}},\ }\bibfield  {title} {\bibinfo {title} {{How well can cold dark matter substructures account for the observed radio flux-ratio anomalies}},\ }\href {https://doi.org/10.1093/mnras/stu2673} {\bibfield  {journal} {\bibinfo  {journal} {Monthly Notices of the Royal Astronomical Society}\ }\textbf {\bibinfo {volume} {447}},\ \bibinfo {pages} {3189} (\bibinfo {year} {2015})},\ \Eprint {https://arxiv.org/abs/1410.3282} {arXiv:1410.3282 [astro-ph.CO]} \BibitemShut {NoStop}%
\bibitem [{\citenamefont {{Hartley}}\ \emph {et~al.}(2019)\citenamefont {{Hartley}}, \citenamefont {{Jackson}}, \citenamefont {{Sluse}}, \citenamefont {{Stacey}},\ and\ \citenamefont {{Vives-Arias}}}]{hartley2019strong}%
  \BibitemOpen
  \bibfield  {author} {\bibinfo {author} {\bibfnamefont {P.}~\bibnamefont {{Hartley}}}, \bibinfo {author} {\bibfnamefont {N.}~\bibnamefont {{Jackson}}}, \bibinfo {author} {\bibfnamefont {D.}~\bibnamefont {{Sluse}}}, \bibinfo {author} {\bibfnamefont {H.~R.}\ \bibnamefont {{Stacey}}},\ and\ \bibinfo {author} {\bibfnamefont {H.}~\bibnamefont {{Vives-Arias}}},\ }\bibfield  {title} {\bibinfo {title} {{Strong lensing reveals jets in a sub-microJy radio-quiet quasar}},\ }\href {https://doi.org/10.1093/mnras/stz510} {\bibfield  {journal} {\bibinfo  {journal} {Monthly Notices of the Royal Astronomical Society}\ }\textbf {\bibinfo {volume} {485}},\ \bibinfo {pages} {3009} (\bibinfo {year} {2019})},\ \Eprint {https://arxiv.org/abs/1901.05791} {arXiv:1901.05791 [astro-ph.GA]} \BibitemShut {NoStop}%
\bibitem [{\citenamefont {{Spingola}}\ \emph {et~al.}(2018)\citenamefont {{Spingola}}, \citenamefont {{McKean}}, \citenamefont {{Auger}}, \citenamefont {{Fassnacht}}, \citenamefont {{Koopmans}}, \citenamefont {{Lagattuta}},\ and\ \citenamefont {{Vegetti}}}]{spingola2018sharp}%
  \BibitemOpen
  \bibfield  {author} {\bibinfo {author} {\bibfnamefont {C.}~\bibnamefont {{Spingola}}}, \bibinfo {author} {\bibfnamefont {J.~P.}\ \bibnamefont {{McKean}}}, \bibinfo {author} {\bibfnamefont {M.~W.}\ \bibnamefont {{Auger}}}, \bibinfo {author} {\bibfnamefont {C.~D.}\ \bibnamefont {{Fassnacht}}}, \bibinfo {author} {\bibfnamefont {L.~V.~E.}\ \bibnamefont {{Koopmans}}}, \bibinfo {author} {\bibfnamefont {D.~J.}\ \bibnamefont {{Lagattuta}}},\ and\ \bibinfo {author} {\bibfnamefont {S.}~\bibnamefont {{Vegetti}}},\ }\bibfield  {title} {\bibinfo {title} {{SHARP - V. Modelling gravitationally lensed radio arcs imaged with global VLBI observations}},\ }\href {https://doi.org/10.1093/mnras/sty1326} {\bibfield  {journal} {\bibinfo  {journal} {Monthly Notices of the Royal Astronomical Society}\ }\textbf {\bibinfo {volume} {478}},\ \bibinfo {pages} {4816} (\bibinfo {year} {2018})},\ \Eprint {https://arxiv.org/abs/1807.05566} {arXiv:1807.05566 [astro-ph.GA]} \BibitemShut {NoStop}%
\bibitem [{\citenamefont {{Biggs}}\ \emph {et~al.}(2004)\citenamefont {{Biggs}}, \citenamefont {{Browne}}, \citenamefont {{Jackson}}, \citenamefont {{York}}, \citenamefont {{Norbury}}, \citenamefont {{McKean}},\ and\ \citenamefont {{Phillips}}}]{biggs2004radio}%
  \BibitemOpen
  \bibfield  {author} {\bibinfo {author} {\bibfnamefont {A.~D.}\ \bibnamefont {{Biggs}}}, \bibinfo {author} {\bibfnamefont {I.~W.~A.}\ \bibnamefont {{Browne}}}, \bibinfo {author} {\bibfnamefont {N.~J.}\ \bibnamefont {{Jackson}}}, \bibinfo {author} {\bibfnamefont {T.}~\bibnamefont {{York}}}, \bibinfo {author} {\bibfnamefont {M.~A.}\ \bibnamefont {{Norbury}}}, \bibinfo {author} {\bibfnamefont {J.~P.}\ \bibnamefont {{McKean}}},\ and\ \bibinfo {author} {\bibfnamefont {P.~M.}\ \bibnamefont {{Phillips}}},\ }\bibfield  {title} {\bibinfo {title} {{Radio, optical and infrared observations of CLASS B0128+437}},\ }\href {https://doi.org/10.1111/j.1365-2966.2004.07701.x} {\bibfield  {journal} {\bibinfo  {journal} {Monthly Notices of the Royal Astronomical Society}\ }\textbf {\bibinfo {volume} {350}},\ \bibinfo {pages} {949} (\bibinfo {year} {2004})},\ \Eprint {https://arxiv.org/abs/astro-ph/0402128} {arXiv:astro-ph/0402128 [astro-ph]} \BibitemShut {NoStop}%
\bibitem [{\citenamefont {{Chan}}\ \emph {et~al.}(2020)\citenamefont {{Chan}}, \citenamefont {{Schive}}, \citenamefont {{Wong}}, \citenamefont {{Chiueh}},\ and\ \citenamefont {{Broadhurst}}}]{chan2020multiple}%
  \BibitemOpen
  \bibfield  {author} {\bibinfo {author} {\bibfnamefont {J.~H.~H.}\ \bibnamefont {{Chan}}}, \bibinfo {author} {\bibfnamefont {H.-Y.}\ \bibnamefont {{Schive}}}, \bibinfo {author} {\bibfnamefont {S.-K.}\ \bibnamefont {{Wong}}}, \bibinfo {author} {\bibfnamefont {T.}~\bibnamefont {{Chiueh}}},\ and\ \bibinfo {author} {\bibfnamefont {T.}~\bibnamefont {{Broadhurst}}},\ }\bibfield  {title} {\bibinfo {title} {{Multiple Images and Flux Ratio Anomaly of Fuzzy Gravitational Lenses}},\ }\href {https://doi.org/10.1103/PhysRevLett.125.111102} {\bibfield  {journal} {\bibinfo  {journal} {Physical Review Letters}\ }\textbf {\bibinfo {volume} {125}},\ \bibinfo {eid} {111102} (\bibinfo {year} {2020})},\ \Eprint {https://arxiv.org/abs/2002.10473} {arXiv:2002.10473 [astro-ph.GA]} \BibitemShut {NoStop}%
\bibitem [{\citenamefont {{Nierenberg}}\ \emph {et~al.}(2014)\citenamefont {{Nierenberg}}, \citenamefont {{Treu}}, \citenamefont {{Wright}}, \citenamefont {{Fassnacht}},\ and\ \citenamefont {{Auger}}}]{nierenberg2014detection}%
  \BibitemOpen
  \bibfield  {author} {\bibinfo {author} {\bibfnamefont {A.~M.}\ \bibnamefont {{Nierenberg}}}, \bibinfo {author} {\bibfnamefont {T.}~\bibnamefont {{Treu}}}, \bibinfo {author} {\bibfnamefont {S.~A.}\ \bibnamefont {{Wright}}}, \bibinfo {author} {\bibfnamefont {C.~D.}\ \bibnamefont {{Fassnacht}}},\ and\ \bibinfo {author} {\bibfnamefont {M.~W.}\ \bibnamefont {{Auger}}},\ }\bibfield  {title} {\bibinfo {title} {{Detection of substructure with adaptive optics integral field spectroscopy of the gravitational lens B1422+231}},\ }\href {https://doi.org/10.1093/mnras/stu862} {\bibfield  {journal} {\bibinfo  {journal} {Monthly Notices of the Royal Astronomical Society}\ }\textbf {\bibinfo {volume} {442}},\ \bibinfo {pages} {2434} (\bibinfo {year} {2014})},\ \Eprint {https://arxiv.org/abs/1402.1496} {arXiv:1402.1496 [astro-ph.GA]} \BibitemShut {NoStop}%
\bibitem [{\citenamefont {{Amara}}\ \emph {et~al.}(2006)\citenamefont {{Amara}}, \citenamefont {{Metcalf}}, \citenamefont {{Cox}},\ and\ \citenamefont {{Ostriker}}}]{amara2006simulations}%
  \BibitemOpen
  \bibfield  {author} {\bibinfo {author} {\bibfnamefont {A.}~\bibnamefont {{Amara}}}, \bibinfo {author} {\bibfnamefont {R.~B.}\ \bibnamefont {{Metcalf}}}, \bibinfo {author} {\bibfnamefont {T.~J.}\ \bibnamefont {{Cox}}},\ and\ \bibinfo {author} {\bibfnamefont {J.~P.}\ \bibnamefont {{Ostriker}}},\ }\bibfield  {title} {\bibinfo {title} {{Simulations of strong gravitational lensing with substructure}},\ }\href {https://doi.org/10.1111/j.1365-2966.2006.10053.x} {\bibfield  {journal} {\bibinfo  {journal} {Monthly Notices of the Royal Astronomical Society}\ }\textbf {\bibinfo {volume} {367}},\ \bibinfo {pages} {1367} (\bibinfo {year} {2006})},\ \Eprint {https://arxiv.org/abs/astro-ph/0411587} {arXiv:astro-ph/0411587 [astro-ph]} \BibitemShut {NoStop}%
\bibitem [{\citenamefont {{Treu}}(2010)}]{treu2010strong}%
  \BibitemOpen
  \bibfield  {author} {\bibinfo {author} {\bibfnamefont {T.}~\bibnamefont {{Treu}}},\ }\bibfield  {title} {\bibinfo {title} {{Strong Lensing by Galaxies}},\ }\href {https://doi.org/10.1146/annurev-astro-081309-130924} {\bibfield  {journal} {\bibinfo  {journal} {Annual Review of Astronomy and Astrophysics}\ }\textbf {\bibinfo {volume} {48}},\ \bibinfo {pages} {87} (\bibinfo {year} {2010})},\ \Eprint {https://arxiv.org/abs/1003.5567} {arXiv:1003.5567 [astro-ph.CO]} \BibitemShut {NoStop}%
\bibitem [{\citenamefont {{Keeton}}\ and\ \citenamefont {{Moustakas}}(2009)}]{keeton2009new}%
  \BibitemOpen
  \bibfield  {author} {\bibinfo {author} {\bibfnamefont {C.~R.}\ \bibnamefont {{Keeton}}}\ and\ \bibinfo {author} {\bibfnamefont {L.~A.}\ \bibnamefont {{Moustakas}}},\ }\bibfield  {title} {\bibinfo {title} {{A New Channel for Detecting Dark Matter Substructure in Galaxies: Gravitational Lens Time Delays}},\ }\href {https://doi.org/10.1088/0004-637X/699/2/1720} {\bibfield  {journal} {\bibinfo  {journal} {The Astrophysical Journal}\ }\textbf {\bibinfo {volume} {699}},\ \bibinfo {pages} {1720} (\bibinfo {year} {2009})},\ \Eprint {https://arxiv.org/abs/0805.0309} {arXiv:0805.0309 [astro-ph]} \BibitemShut {NoStop}%
\bibitem [{\citenamefont {{Schneider}}\ and\ \citenamefont {{Sluse}}(2014)}]{schneider2013source}%
  \BibitemOpen
  \bibfield  {author} {\bibinfo {author} {\bibfnamefont {P.}~\bibnamefont {{Schneider}}}\ and\ \bibinfo {author} {\bibfnamefont {D.}~\bibnamefont {{Sluse}}},\ }\bibfield  {title} {\bibinfo {title} {{Source-position transformation: an approximate invariance in strong gravitational lensing}},\ }\href {https://doi.org/10.1051/0004-6361/201322106} {\bibfield  {journal} {\bibinfo  {journal} {Astronomy \& Astrophysics}\ }\textbf {\bibinfo {volume} {564}},\ \bibinfo {eid} {A103} (\bibinfo {year} {2014})},\ \Eprint {https://arxiv.org/abs/1306.4675} {arXiv:1306.4675 [astro-ph.CO]} \BibitemShut {NoStop}%
\bibitem [{\citenamefont {{Planck Collaboration}}(2020)}]{aghanim2020planck}%
  \BibitemOpen
  \bibfield  {author} {\bibinfo {author} {\bibnamefont {{Planck Collaboration}}},\ }\bibfield  {title} {\bibinfo {title} {{Planck 2018 results. VI. Cosmological parameters}},\ }\href {https://doi.org/10.1051/0004-6361/201833910} {\bibfield  {journal} {\bibinfo  {journal} {Astronomy \& Astrophysics}\ }\textbf {\bibinfo {volume} {641}},\ \bibinfo {eid} {A6} (\bibinfo {year} {2020})},\ \Eprint {https://arxiv.org/abs/1807.06209} {arXiv:1807.06209 [astro-ph.CO]} \BibitemShut {NoStop}%
\bibitem [{\citenamefont {{Kochanek}}\ \emph {et~al.}(1999)\citenamefont {{Kochanek}}, \citenamefont {{Falco}}, \citenamefont {{Impey}}, \citenamefont {{Leh{\'a}r}}, \citenamefont {{McLeod}},\ and\ \citenamefont {{Rix}}}]{kochanek1999results}%
  \BibitemOpen
  \bibfield  {author} {\bibinfo {author} {\bibfnamefont {C.~S.}\ \bibnamefont {{Kochanek}}}, \bibinfo {author} {\bibfnamefont {E.~E.}\ \bibnamefont {{Falco}}}, \bibinfo {author} {\bibfnamefont {C.~D.}\ \bibnamefont {{Impey}}}, \bibinfo {author} {\bibfnamefont {J.}~\bibnamefont {{Leh{\'a}r}}}, \bibinfo {author} {\bibfnamefont {B.~A.}\ \bibnamefont {{McLeod}}},\ and\ \bibinfo {author} {\bibfnamefont {H.~W.}\ \bibnamefont {{Rix}}},\ }\bibfield  {title} {\bibinfo {title} {{Results from the CASTLES survey of gravitational lenses}},\ }in\ \href {https://doi.org/10.1063/1.58598} {\emph {\bibinfo {booktitle} {After the Dark Ages: When Galaxies were Young (the Universe at 2 < Z < 5)}}},\ \bibinfo {series} {American Institute of Physics Conference Series}, Vol.\ \bibinfo {volume} {470},\ \bibinfo {editor} {edited by\ \bibinfo {editor} {\bibfnamefont {S.}~\bibnamefont {{Holt}}}\ and\ \bibinfo {editor} {\bibfnamefont {E.}~\bibnamefont {{Smith}}}}\ (\bibinfo  {publisher} {AIP},\ \bibinfo {year} {1999})\ pp.\ \bibinfo
  {pages} {163--175},\ \Eprint {https://arxiv.org/abs/astro-ph/9811111} {arXiv:astro-ph/9811111 [astro-ph]} \BibitemShut {NoStop}%
\bibitem [{\citenamefont {Lindeberg}(1922)}]{lindeberg1922neue}%
  \BibitemOpen
  \bibfield  {author} {\bibinfo {author} {\bibfnamefont {J.~W.}\ \bibnamefont {Lindeberg}},\ }\bibfield  {title} {\bibinfo {title} {Eine neue herleitung des exponentialgesetzes in der wahrscheinlichkeitsrechnung},\ }\href {https://link.springer.com/article/10.1007/BF01494395} {\bibfield  {journal} {\bibinfo  {journal} {Mathematische Zeitschrift}\ }\textbf {\bibinfo {volume} {15}},\ \bibinfo {pages} {211} (\bibinfo {year} {1922})}\BibitemShut {NoStop}%
\bibitem [{\citenamefont {Cam}(1986)}]{CLT}%
  \BibitemOpen
  \bibfield  {author} {\bibinfo {author} {\bibfnamefont {L.~L.}\ \bibnamefont {Cam}},\ }\bibfield  {title} {\bibinfo {title} {The central limit theorem around 1935},\ }\href {http://www.jstor.org/stable/2245503} {\bibfield  {journal} {\bibinfo  {journal} {Statistical Science}\ }\textbf {\bibinfo {volume} {1}},\ \bibinfo {pages} {78} (\bibinfo {year} {1986})}\BibitemShut {NoStop}%
\bibitem [{\citenamefont {{Murray}}(2018)}]{murray2018powerbox}%
  \BibitemOpen
  \bibfield  {author} {\bibinfo {author} {\bibfnamefont {S.~G.}\ \bibnamefont {{Murray}}},\ }\bibfield  {title} {\bibinfo {title} {{powerbox: A Python package for creating structured fields with isotropic power spectra}},\ }\href {https://doi.org/10.21105/joss.00850} {\bibfield  {journal} {\bibinfo  {journal} {The Journal of Open Source Software}\ }\textbf {\bibinfo {volume} {3}},\ \bibinfo {eid} {850} (\bibinfo {year} {2018})},\ \Eprint {https://arxiv.org/abs/1809.05030} {arXiv:1809.05030 [astro-ph.IM]} \BibitemShut {NoStop}%
\bibitem [{\citenamefont {{Gao}}\ \emph {et~al.}(2022)\citenamefont {{Gao}}, \citenamefont {{Li}},\ and\ \citenamefont {{Gao}}}]{gao2022prospects}%
  \BibitemOpen
  \bibfield  {author} {\bibinfo {author} {\bibfnamefont {R.}~\bibnamefont {{Gao}}}, \bibinfo {author} {\bibfnamefont {Z.}~\bibnamefont {{Li}}},\ and\ \bibinfo {author} {\bibfnamefont {H.}~\bibnamefont {{Gao}}},\ }\bibfield  {title} {\bibinfo {title} {{Prospects of strongly lensed fast radio bursts: simultaneous measurement of post-Newtonian parameter and Hubble constant}},\ }\href {https://doi.org/10.1093/mnras/stac2270} {\bibfield  {journal} {\bibinfo  {journal} {Monthly Notices of the Royal Astronomical Society}\ }\textbf {\bibinfo {volume} {516}},\ \bibinfo {pages} {1977} (\bibinfo {year} {2022})},\ \Eprint {https://arxiv.org/abs/2208.10175} {arXiv:2208.10175 [astro-ph.CO]} \BibitemShut {NoStop}%
\bibitem [{\citenamefont {{Saha}}\ and\ \citenamefont {{Williams}}(2003)}]{saha2003qualitative}%
  \BibitemOpen
  \bibfield  {author} {\bibinfo {author} {\bibfnamefont {P.}~\bibnamefont {{Saha}}}\ and\ \bibinfo {author} {\bibfnamefont {L.~L.~R.}\ \bibnamefont {{Williams}}},\ }\bibfield  {title} {\bibinfo {title} {{Qualitative Theory for Lensed QSOs}},\ }\href {https://doi.org/10.1086/375204} {\bibfield  {journal} {\bibinfo  {journal} {The Astronomical Journal}\ }\textbf {\bibinfo {volume} {125}},\ \bibinfo {pages} {2769} (\bibinfo {year} {2003})},\ \Eprint {https://arxiv.org/abs/astro-ph/0303261} {arXiv:astro-ph/0303261 [astro-ph]} \BibitemShut {NoStop}%
\bibitem [{\citenamefont {{Oguri}}(2019)}]{2019RPPh...82l6901O}%
  \BibitemOpen
  \bibfield  {author} {\bibinfo {author} {\bibfnamefont {M.}~\bibnamefont {{Oguri}}},\ }\bibfield  {title} {\bibinfo {title} {{Strong gravitational lensing of explosive transients}},\ }\href {https://doi.org/10.1088/1361-6633/ab4fc5} {\bibfield  {journal} {\bibinfo  {journal} {Reports on Progress in Physics}\ }\textbf {\bibinfo {volume} {82}},\ \bibinfo {eid} {126901} (\bibinfo {year} {2019})},\ \Eprint {https://arxiv.org/abs/1907.06830} {arXiv:1907.06830 [astro-ph.CO]} \BibitemShut {NoStop}%
\bibitem [{\citenamefont {{Liao}}\ \emph {et~al.}(2022)\citenamefont {{Liao}}, \citenamefont {{Biesiada}},\ and\ \citenamefont {{Zhu}}}]{2022ChPhL..39k9801L}%
  \BibitemOpen
  \bibfield  {author} {\bibinfo {author} {\bibfnamefont {K.}~\bibnamefont {{Liao}}}, \bibinfo {author} {\bibfnamefont {M.}~\bibnamefont {{Biesiada}}},\ and\ \bibinfo {author} {\bibfnamefont {Z.-H.}\ \bibnamefont {{Zhu}}},\ }\bibfield  {title} {\bibinfo {title} {{Strongly Lensed Transient Sources: A Review}},\ }\href {https://doi.org/10.1088/0256-307X/39/11/119801} {\bibfield  {journal} {\bibinfo  {journal} {Chinese Physics Letters}\ }\textbf {\bibinfo {volume} {39}},\ \bibinfo {eid} {119801} (\bibinfo {year} {2022})},\ \Eprint {https://arxiv.org/abs/2207.13489} {arXiv:2207.13489 [astro-ph.HE]} \BibitemShut {NoStop}%
\bibitem [{\citenamefont {{Suyu}}\ \emph {et~al.}(2024)\citenamefont {{Suyu}}, \citenamefont {{Goobar}}, \citenamefont {{Collett}}, \citenamefont {{More}},\ and\ \citenamefont {{Vernardos}}}]{2024SSRv..220...13S}%
  \BibitemOpen
  \bibfield  {author} {\bibinfo {author} {\bibfnamefont {S.~H.}\ \bibnamefont {{Suyu}}}, \bibinfo {author} {\bibfnamefont {A.}~\bibnamefont {{Goobar}}}, \bibinfo {author} {\bibfnamefont {T.}~\bibnamefont {{Collett}}}, \bibinfo {author} {\bibfnamefont {A.}~\bibnamefont {{More}}},\ and\ \bibinfo {author} {\bibfnamefont {G.}~\bibnamefont {{Vernardos}}},\ }\bibfield  {title} {\bibinfo {title} {{Strong Gravitational Lensing and Microlensing of Supernovae}},\ }\href {https://doi.org/10.1007/s11214-024-01044-7} {\bibfield  {journal} {\bibinfo  {journal} {Space Science Reviews}\ }\textbf {\bibinfo {volume} {220}},\ \bibinfo {eid} {13} (\bibinfo {year} {2024})},\ \Eprint {https://arxiv.org/abs/2301.07729} {arXiv:2301.07729 [astro-ph.CO]} \BibitemShut {NoStop}%
\bibitem [{\citenamefont {{Liao}}\ \emph {et~al.}(2015)\citenamefont {{Liao}}, \citenamefont {{Treu}}, \citenamefont {{Marshall}},\ and\ \citenamefont {{et al.}}}]{2015ApJ...800...11L}%
  \BibitemOpen
  \bibfield  {author} {\bibinfo {author} {\bibfnamefont {K.}~\bibnamefont {{Liao}}}, \bibinfo {author} {\bibfnamefont {T.}~\bibnamefont {{Treu}}}, \bibinfo {author} {\bibfnamefont {P.}~\bibnamefont {{Marshall}}},\ and\ \bibinfo {author} {\bibnamefont {{et al.}}},\ }\bibfield  {title} {\bibinfo {title} {{Strong Lens Time Delay Challenge. II. Results of TDC1}},\ }\href {https://doi.org/10.1088/0004-637X/800/1/11} {\bibfield  {journal} {\bibinfo  {journal} {The Astrophysical Journal}\ }\textbf {\bibinfo {volume} {800}},\ \bibinfo {eid} {11} (\bibinfo {year} {2015})},\ \Eprint {https://arxiv.org/abs/1409.1254} {arXiv:1409.1254 [astro-ph.IM]} \BibitemShut {NoStop}%
\bibitem [{\citenamefont {{Li}}\ \emph {et~al.}(2018)\citenamefont {{Li}}, \citenamefont {{Gao}}, \citenamefont {{Ding}}, \citenamefont {{Wang}},\ and\ \citenamefont {{Zhang}}}]{2018NatCo...9.3833L}%
  \BibitemOpen
  \bibfield  {author} {\bibinfo {author} {\bibfnamefont {Z.-X.}\ \bibnamefont {{Li}}}, \bibinfo {author} {\bibfnamefont {H.}~\bibnamefont {{Gao}}}, \bibinfo {author} {\bibfnamefont {X.-H.}\ \bibnamefont {{Ding}}}, \bibinfo {author} {\bibfnamefont {G.-J.}\ \bibnamefont {{Wang}}},\ and\ \bibinfo {author} {\bibfnamefont {B.}~\bibnamefont {{Zhang}}},\ }\bibfield  {title} {\bibinfo {title} {{Strongly lensed repeating fast radio bursts as precision probes of the universe}},\ }\href {https://doi.org/10.1038/s41467-018-06303-0} {\bibfield  {journal} {\bibinfo  {journal} {Nature Communications}\ }\textbf {\bibinfo {volume} {9}},\ \bibinfo {eid} {3833} (\bibinfo {year} {2018})},\ \Eprint {https://arxiv.org/abs/1708.06357} {arXiv:1708.06357 [astro-ph.CO]} \BibitemShut {NoStop}%
\bibitem [{\citenamefont {{Liao}}\ \emph {et~al.}(2017)\citenamefont {{Liao}}, \citenamefont {{Fan}}, \citenamefont {{Ding}}, \citenamefont {{Biesiada}},\ and\ \citenamefont {{Zhu}}}]{liao2017precision}%
  \BibitemOpen
  \bibfield  {author} {\bibinfo {author} {\bibfnamefont {K.}~\bibnamefont {{Liao}}}, \bibinfo {author} {\bibfnamefont {X.-L.}\ \bibnamefont {{Fan}}}, \bibinfo {author} {\bibfnamefont {X.}~\bibnamefont {{Ding}}}, \bibinfo {author} {\bibfnamefont {M.}~\bibnamefont {{Biesiada}}},\ and\ \bibinfo {author} {\bibfnamefont {Z.-H.}\ \bibnamefont {{Zhu}}},\ }\bibfield  {title} {\bibinfo {title} {{Precision cosmology from future lensed gravitational wave and electromagnetic signals}},\ }\href {https://doi.org/10.1038/s41467-017-01152-9} {\bibfield  {journal} {\bibinfo  {journal} {Nature Communications}\ }\textbf {\bibinfo {volume} {8}},\ \bibinfo {eid} {1148} (\bibinfo {year} {2017})},\ \Eprint {https://arxiv.org/abs/1703.04151} {arXiv:1703.04151 [astro-ph.CO]} \BibitemShut {NoStop}%
\bibitem [{\citenamefont {{Gilman}}\ \emph {et~al.}(2020)\citenamefont {{Gilman}}, \citenamefont {{Birrer}}, \citenamefont {{Nierenberg}}, \citenamefont {{Treu}}, \citenamefont {{Du}},\ and\ \citenamefont {{Benson}}}]{gilman2020warm}%
  \BibitemOpen
  \bibfield  {author} {\bibinfo {author} {\bibfnamefont {D.}~\bibnamefont {{Gilman}}}, \bibinfo {author} {\bibfnamefont {S.}~\bibnamefont {{Birrer}}}, \bibinfo {author} {\bibfnamefont {A.}~\bibnamefont {{Nierenberg}}}, \bibinfo {author} {\bibfnamefont {T.}~\bibnamefont {{Treu}}}, \bibinfo {author} {\bibfnamefont {X.}~\bibnamefont {{Du}}},\ and\ \bibinfo {author} {\bibfnamefont {A.}~\bibnamefont {{Benson}}},\ }\bibfield  {title} {\bibinfo {title} {{Warm dark matter chills out: constraints on the halo mass function and the free-streaming length of dark matter with eight quadruple-image strong gravitational lenses}},\ }\href {https://doi.org/10.1093/mnras/stz3480} {\bibfield  {journal} {\bibinfo  {journal} {Monthly Notices of the Royal Astronomical Society}\ }\textbf {\bibinfo {volume} {491}},\ \bibinfo {pages} {6077} (\bibinfo {year} {2020})},\ \Eprint {https://arxiv.org/abs/1908.06983} {arXiv:1908.06983 [astro-ph.CO]} \BibitemShut {NoStop}%
\bibitem [{\citenamefont {{Dike}}\ \emph {et~al.}(2023)\citenamefont {{Dike}}, \citenamefont {{Gilman}},\ and\ \citenamefont {{Treu}}}]{dike2023strong}%
  \BibitemOpen
  \bibfield  {author} {\bibinfo {author} {\bibfnamefont {V.}~\bibnamefont {{Dike}}}, \bibinfo {author} {\bibfnamefont {D.}~\bibnamefont {{Gilman}}},\ and\ \bibinfo {author} {\bibfnamefont {T.}~\bibnamefont {{Treu}}},\ }\bibfield  {title} {\bibinfo {title} {{Strong lensing constraints on primordial black holes as a dark matter candidate}},\ }\href {https://doi.org/10.1093/mnras/stad1313} {\bibfield  {journal} {\bibinfo  {journal} {Monthly Notices of the Royal Astronomical Society}\ }\textbf {\bibinfo {volume} {522}},\ \bibinfo {pages} {5434} (\bibinfo {year} {2023})},\ \Eprint {https://arxiv.org/abs/2210.09493} {arXiv:2210.09493 [astro-ph.CO]} \BibitemShut {NoStop}%
\bibitem [{\citenamefont {{Laroche}}\ \emph {et~al.}(2022)\citenamefont {{Laroche}}, \citenamefont {{Gilman}}, \citenamefont {{Li}}, \citenamefont {{Bovy}},\ and\ \citenamefont {{Du}}}]{laroche2022quantum}%
  \BibitemOpen
  \bibfield  {author} {\bibinfo {author} {\bibfnamefont {A.}~\bibnamefont {{Laroche}}}, \bibinfo {author} {\bibfnamefont {D.}~\bibnamefont {{Gilman}}}, \bibinfo {author} {\bibfnamefont {X.}~\bibnamefont {{Li}}}, \bibinfo {author} {\bibfnamefont {J.}~\bibnamefont {{Bovy}}},\ and\ \bibinfo {author} {\bibfnamefont {X.}~\bibnamefont {{Du}}},\ }\bibfield  {title} {\bibinfo {title} {{Quantum fluctuations masquerade as haloes: bounds on ultra-light dark matter from quadruply imaged quasars}},\ }\href {https://doi.org/10.1093/mnras/stac2677} {\bibfield  {journal} {\bibinfo  {journal} {Monthly Notices of the Royal Astronomical Society}\ }\textbf {\bibinfo {volume} {517}},\ \bibinfo {pages} {1867} (\bibinfo {year} {2022})},\ \Eprint {https://arxiv.org/abs/2206.11269} {arXiv:2206.11269 [astro-ph.CO]} \BibitemShut {NoStop}%
\end{thebibliography}%

\end{document}